\documentclass[a4paper,11pt]{article}
\pdfoutput=1 

\usepackage{jheppub,cancel}
\usepackage[T1]{fontenc}
\usepackage{xcolor}
\usepackage{subfig}
\usepackage{slashed}
\usepackage{multirow}

\newcommand{\bea}{\begin{eqnarray}}  
\newcommand{\eea}{\end{eqnarray}}  

\newcommand{\ba}{\begin{array}}
\newcommand{\ea}{\end{array}}

\title{New leptons with exotic decays: collider limits and
  dark matter complementarity}

\author[a,b]{Guilherme Guedes,}
\affiliation[a]{Laborat\'orio de Instrumenta\c cao e F\'isica
  Experimental de Part\'iculas, Departamento de 
F\'isica da Universidade do Minho, Campus de Gualtar, 4710-057 Braga,
Portugal} 
\emailAdd{gguedes@lip.pt}
\author[b]{and Jos\'e Santiago}

\affiliation[b]{CAFPE and Departamento de F{\'\i}sica Te{\'o}rica y
  del Cosmos, Universidad 
de Granada, E\textendash{}18071 Granada, Spain}

\emailAdd{jsantiago@ugr.es}

\abstract{
We describe current and future hadron collider limits on new
vector-like leptons with exotic decays. We consider the possibility
that, besides standard decays, the new leptons can also decay into a
Standard Model charged lepton and a stable particle like a dark
photon. To increase their applicability, our results are given in
terms of arbitrary branching 
ratios in the different decay channels. In the case that the dark
photon is stable at cosmological scales we discuss the
interplay between the dark photon and the vector-like lepton in
generating the observed dark matter relic abundance and the
complementarity of collider searches and dark matter phenomenology.
}

\begin{document}
	\maketitle
	\flushbottom

\section{Introduction \label{Introduction}}

New fermions are a common
occurrence in models of physics beyond the Standard Model (SM). 
If they are vector-like~\cite{delAguila:1982fs}, namely both
chiralities have the same quantum 
numbers, their mass term is gauge invariant and therefore it is not
tied to the electroweak scale. As a result, they do not contribute to
anomalies and all their physical effects decouple as inverse powers of
their mass. Their phenomenological implications
have been extensively studied, in particular in the
case of vector-like quarks (triplets under color $SU(3)_C$), as they
are strongly pair-produced at hadron colliders. Furthermore,
the fact that electroweak top couplings have been measured with
less accuracy than for lighter fermions leaves more room for
relatively large indirect
effects~\cite{delAguila:2000rc,delAguila:2008pw} and single 
production (see
however~\cite{Aguilar-Saavedra:2013qpa} for strong constraints in
minimal models and~\cite{Anastasiou:2009rv} 
for ways to evade them in more realistic ones).

Vector-like leptons (VLL), neutral under $SU(3)_C$, 
have received much less attention. Indirect
constraints~\cite{deBlas:2013gla} put
very stringent limits on their mixing with the SM fermions, thus
significantly reducing the possibility of a sizeable single production
at colliders.  Pair-production via Drell-Yan is quite model
independent (see however~\cite{Araque:2015cna,Araque:2015cbe,Chala:2020odv}) but the smaller
production cross-section than for vector-like quarks makes the reach
quite modest
(see~\cite{Altmannshofer:2013zba,Falkowski:2013jya,Dermisek:2014qca,Kumar:2015tna,Bhattiprolu:2019vdu} 
for theoretical studies and~\cite{Aaboud:2018jiw,Aad:2015dha,Sirunyan:2019ofn} for experimental searches). 
Furthermore, in all these cases, decays into only SM
particles are 
assumed.  However, there are classes of models with new VLL that
incorporate a discrete symmetry under which SM particles are even and
new particles are odd, thus preventing the decay of the VLL into only
SM particles. They typically decay into the lightest odd-symmetric 
particle, which is often a dark matter (DM) candidate. A prime example
is T-parity in Little Higgs models~\cite{Cheng:2004yc,Low:2004xc}. The
lightest (and 
therefore easiest to produce) VLL usually decays into a SM lepton and a
stable particle that results in missing energy at colliders. Such a
decay has not been considered by experimental collaborations in the
context of VLL searches. The production and decay pattern is very
similar to the one of slepton pair production but due to the different
spin of the particles involved, the interpretation of the experimental
results in terms of VLL searches requires a recast of the analysis by
theorists (see for instance~\cite{Dercks:2018hgz}).

Even more interestingly, the possibility of simultaneously having both types of 
decays, into a SM lepton plus a $W$, $Z$ or Higgs boson and into a SM
lepton and missing energy, has never been considered in the past,
despite the fact that this possibility is easy to realize and is even
well motivated in the context of feebly interacting
DM~\cite{Delaunay:2020vdb}. In this article we consider the 
possibility that the new VLL can simultaneously decay into the usual
SM final states as well as into a SM lepton and missing energy. We
will leave the decay pattern completely general so that our results
apply to a large number of phenomenological models involving
VLL. (See~\cite{Chala:2017xgc} for a similar study for the
case of vector-like quarks.)

Inspired by the case of Little Higgs models with T-parity and by
feebly interacting dark photon models we will consider the missing
energy particle to be a dark photon, a massive vector that is stable
at detector scales. However, this dark photon could be stable at much
longer scales, of the order of the lifetime of the Universe and
therefore be a good DM candidate. We will also explore
this possibility and we will show that the VLL can play a crucial role
in this regard. Indeed, it can open a large region of the allowed
parameter space by either contributing to the relic abundance via
co-annihilation with the dark photon or via the freeze-in mechanism. We
will analyze these two possibilities and we will show that they can
give complementary information in the former case and benefit from the
collider searches in the latter one.

The rest of this article is organized as follows. We describe in
Section~\ref{searches} the most relevant current experimental searches
for a new VLL with general decays. We then optimize these searches and
obtain the expected LHC bounds on new VLL with arbitrary branching
ratios with the current recorded luminosity. This is one of the main
results of this article and it allows us to immediately get the
constraints on new VLL with arbitrary decay patterns. We then explore
the reach of the high-luminosity (HL-LHC) and high-energy (HE-LHC)
configurations of the LHC together with an estimation of the final
hh-FCC reach. Section \ref{sec:dm} is devoted to the case in which the
missing energy particle is stable and can act as a good DM candidate,
first assuming the standard freeze-out mechanism and then the
freeze-in one. We will see that in both cases the interplay with the
VLL is crucial for a successful model. We then present our conclusions
in Section~\ref{conclusions}. We present in Appendix~\ref{realization}
an explicit realization of the scenario we consider in the main text.

\section{New vector-like leptons with general decays\label{searches}}

The goal of this article is to study the current and future reach of
hadron colliders on new VLL that can decay not only into SM particles
but also into a SM charged lepton and a neutral particle that is
stable at detector scales and therefore appears as missing energy. We
will present our results in a model-independent way whenever possible,
as a function of arbitrary branching ratios in the different
channels. To show actual limits we will however focus on a new VLL
singlet with electric charge -1, $E_{L,R}$, and mass $M_E$, and a
massive vector boson 
$A_H^\mu$, as the stable (at detector scales) particle, with mass
$M_{A_{H}}<M_E$ so that $E$ can decay into $A_H$ and a SM lepton.

An explicit
realization of our model is given in Appendix~\ref{realization} but
the details are not needed for the moment.
The only relevant information is that 
the dominant $E$ decays are given by the
following branching ratios $\mathrm{BR}(E\to \ell H)$,
$\mathrm{BR}(E\to \nu_{\ell} W)$, $\mathrm{BR}(E\to\ell Z)$  
and $\mathrm{BR}(E\to \ell A_H)$, where $\ell$ stands for either
electron or muon\footnote{Decays into tau leptons have been considered,
assuming SM decays only,
in~\cite{delAguila:2010es,Kumar:2015tna,Bhattiprolu:2019vdu}.} 
and we assume the sum of these four branching ratios to be equal to
one but otherwise arbitrary.\footnote{The decays into SM particles are
usually fixed by the quantum numbers of the VLL but in realistic
models with a rich spectrum, the mixing between heavy states can lead
to arbitrary decay patterns~\cite{Chala:2013ega}.}  We focus on $E$
Drell-Yan pair production, with 
subsequent decays governed by the corresponding branching
ratios.\footnote{Studies in which the production and/or decay of new
vector-like fermions are dominated by non-renormalizable interactions
can be found in~\cite{Criado:2019mvu,Chala:2020odv}.} 
Out of the four possible decay channels, the two that are
easiest to detect experimentally are $E\to \ell Z$ and
$E\to \ell A_H$. The charged current one into
$\nu_\ell W$
is difficult to disentangle from the overwhelming
$W+\mbox{jets}$ background and the one into $\ell H$ is either also
difficult to disentangle from the relevant background or suffers from
small branching fractions into easier to detect channels.
Thus, in the following we will focus on the cleaner
channels and give our results in terms of $\mathrm{BR}(E\to \ell Z)$ and $\mathrm{BR}(E
\to \ell A_H)$.
We will show that the
results are mostly insensitive to the value of the two extra
branching ratios. 

There are currently two 
experimental analyses that are most sensitive to these discovery
channels, searches for VLL into $\ell Z$ and slepton searches. Neither
of them can be directly used in our more general scenario, except for
the former in the $\mathrm{BR}(E\to \ell A_H)=0$ limit. The slepton searches
have to be completely recast because of the different spin of the
intermediate particle and also because of the contamination of other
channels in the  
$\mathrm{BR}(E\to \ell A_H)\neq 1$ limit.

We will begin this section by recasting the two relevant experimental
analyses. We will first compare our results with the ones published by
the experimental collaborations and then extend the analyses by
considering arbitrary decays into the different channels. We will also
update the analyses to take full advantage of higher luminosity and/or
center of mass energy. 

\subsection{Recasting existing analyses}

Since our goal is to interpolate between the limiting cases in which
the branching ratio of the VLL to the missing energy channel goes from
0 to 1, we start by reproducing searches that probe these two
limiting cases. The VLL model is implemented in
\texttt{Feynrules}~\cite{Alloul:2013bka} 
 and leading order event generation is done with
 \texttt{MadGraph5\_aMC@NLO} \cite{Alwall:2014hca}. For the background
 simulation generator level cuts are applied which are specified in
 the text. All 
   of these were tested to verify that their influence was minimal to
   the final yield of events after the analysis. Showering and 
 hadronization are performed by \texttt{Pythia8}~\cite{Sjostrand:2014zea} 
 with the CMS CUETP8M1~\cite{Khachatryan:2015pea}
 underlying event tuning and the \texttt{NNPDF23LO} \cite{Ball:2012cx}
 parton distribution functions. The detector response is modeled with
 \texttt{Delphes 3}~\cite{deFavereau:2013fsa}. We use the default CMS card for the LHC analysis and the HL-LHC detector card for the $\sqrt{s} = 27\,\mathrm{TeV}$ analysis.
$95\%$ C.L. limits are obtained using the $CL_s$ \cite{Read:2002hq} 
method by fitting the relevant discriminant variables using \texttt{OpTHyLic}
\cite{Busato:2015ola} 
 which outputs the upper limit on the signal strength, $\mu =
 \sigma_{up}/ \sigma_{th}$, where $\sigma_{up}$ is the upper limit on
 the cross-section and $\sigma_{th}$ is the theoretical prediction
 obtained through the MadGraph simulation. 

 \begin{figure}[t]
\centering
  \includegraphics[width=0.5\linewidth]{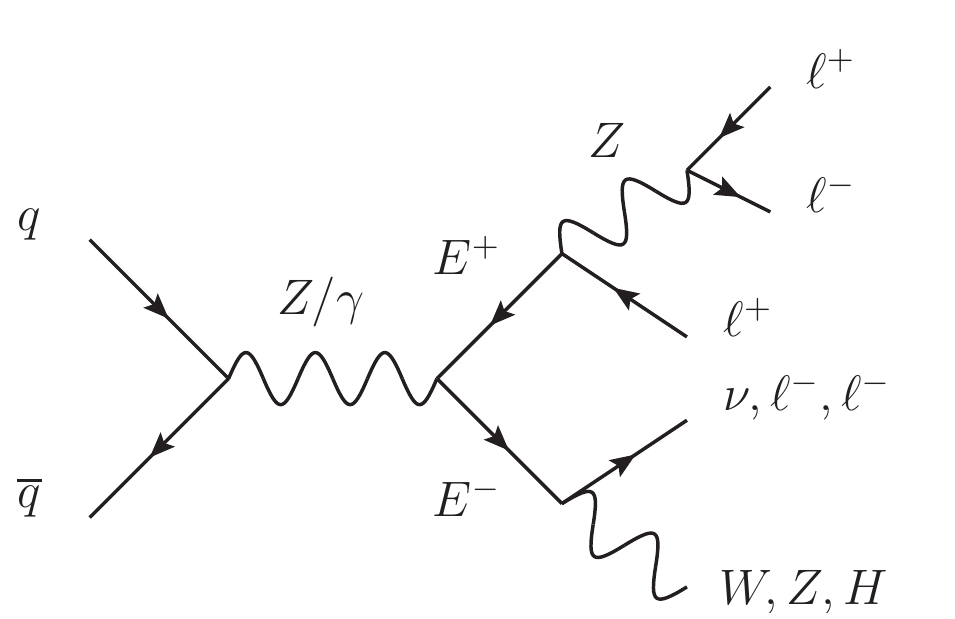}
  \caption{Pair production of a VLL singlet $E$ and decay channels our
    analysis is most sensitive to.}
  \label{fig:zlprocess}
\end{figure}

\subsubsection{Decays into SM particles}

For the case in which the VLL decays exclusively to SM final states
($W\nu$, $Z\ell$ and $H\ell$) we reproduce the analysis presented in
Ref. \cite{Aad:2015dha}, an ATLAS search performed at
$\sqrt{s}=8\;\mathrm{TeV}$ and an integrated luminosity of
$\mathcal{L} = 20.3\; \mathrm{fb}^{-1}$, looking for multi-lepton
signals coming from the $Z \ell$ decay of a singlet VLL. The
main production and decay channels are depicted
in Figure \ref{fig:zlprocess}. This analysis selects 2 opposite sign
same flavour (OSSF) leptons to reconstruct a $Z$ boson and a third
lepton with a $\Delta R\equiv \sqrt{\Delta\eta^2+\Delta \phi^2} < 3$,
with $\eta$ and $\phi$ the pseudo rapidity and azimuthal angle,
respectively, from the reconstructed $Z$ boson, which
is defined as the off-$Z$ lepton. The definition of the full cuts and the
corresponding efficiencies are presented in Table
\ref{VLL:cuts:table}.
\begin{table}[b]
\begin{center}
 \begin{tabular}{|c c c c |} 
 \hline
 Selection Cuts &
 $ZZ$
 & $WZ$
 &
 $Z\gamma$ 
 \\
\hline\hline
\multicolumn{1}{|c}{\begin{tabular}[c]{@{}c@{}}OSSF lepton pair with\\$|m_{\ell\ell}- m_Z| < 10\;\mathrm{GeV}$\end{tabular}}& 0.25 & 0.19 & 0.0024\\ 
 \hline
 $p^{\ell_1}_T > 26\;\mathrm{GeV}$ & 0.25 & 0.19 & 0.0023\\
 \hline
 $\Delta R (Z,\mathrm{off-}Z\;\mathrm{lepton}) < 3$ & 0.17 & 0.11 & 0.0008 \\
 \hline
\end{tabular}
\caption{\label{VLL:cuts:table} Cumulative efficiencies for the
  background events 
  after applying the selection cuts corresponding to the analysis
  performed at $\sqrt{s} = 8 \;\mathrm{TeV}$. Efficiencies are
  presented as the number of events which are selected over the number
  of initial events.
  We have included in the generation only leptonic
  decays of the $Z$ and $W$ into electron, muons and taus but apply
  our cuts only to final state electrons and muons.
  } 
\end{center}
\end{table}
The analysis searches for an excess in the
distribution of the variable $\Delta m = m_{3\ell} - m_{\ell\ell}$,
where the mass of the reconstructed $Z$ boson, denoted by
$m_{\ell\ell}$, is subtracted from the 
invariant mass of the 3-lepton system, $m_{3\ell}$.
Furthermore, 3 exclusive signal regions are defined, depending on the
number of identified leptons and hadronically decaying $W$: 4-lepton
region, in which at least 4 leptons are identified; 3-lepton + $jj$
region, in which precisely 3 leptons are identified together with 2
jets whose invariant mass must be in the range
  $m_W
  -20\;\mathrm{GeV} < m_{jj} < 150 \;\mathrm{GeV}$ , with $m_W$
  the $W$ boson mass; and 3-lepton only region, in which exactly 3
  leptons are identified with no pairs of jets satisfying the previous
  condition on their invariant mass.

\begin{figure}[t]%
    \centering
    \includegraphics[width=0.6\linewidth]{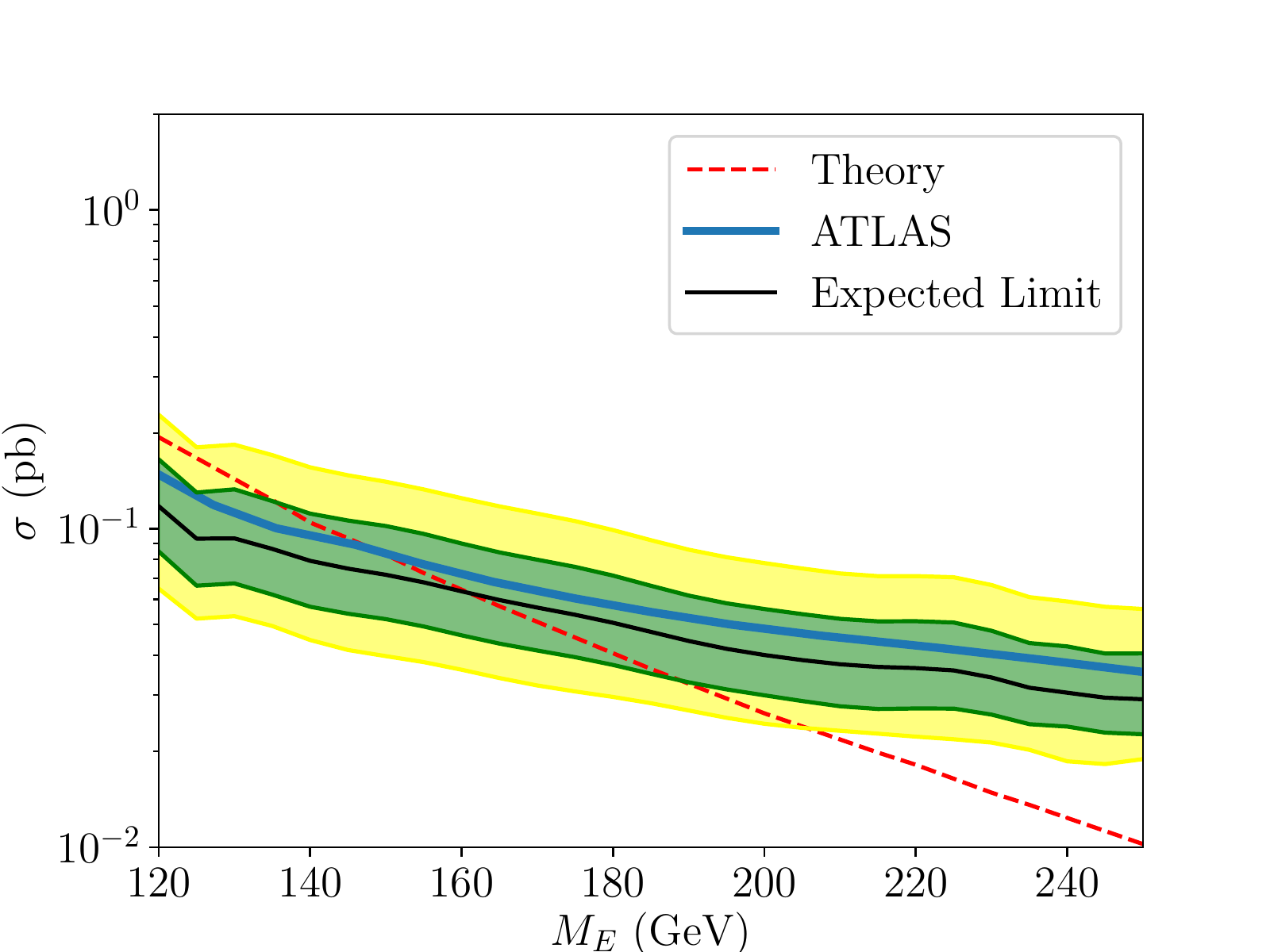}
    \caption{Comparison of our recast of the VLL search with the ATLAS
      collaboration results. We show the case in which the off-$Z$
      lepton is an electron, with the 1 
      (green) and 2 (yellow) sigma exclusion region from our
      simulation together with the expected limit as reported
      in the ATLAS search (solid blue) and the theoretical pair
      production cross section (dashed red). The branching fractions
      are fixed to those of a VLL singlet with SM only decays (as a
      function of its mass).}
    \label{fig:comparison}%
\end{figure}

The analysis is performed separately for the case in which the off-$Z$
lepton is an electron or a muon, corresponding to the VLL coupling
only to first or second generation leptons, respectively. The main
backgrounds for this 
analysis are $ZZ$, $WZ$ and $Z\gamma$, for which our simulation very
accurately reproduces the shape. We normalize these backgrounds to the
values reported in the experimental publication, which amounts to a
factor between ~1.4 and 3.5, depending on the signal region, including
the corresponding K-factor. 
 We show in
Figure \ref{fig:comparison} the comparison of our 1- and 2-sigma
exclusion plot (Brazilian plot) with the expected limit reported in the
experimental search, together with the theoretical pair production of
the VLL, for the case in which the off-$Z$ lepton is an electron. The case in which it is a muon shows a similar level
of agreement. In this analysis the VLL branching fractions are fixed
to those of an electroweak singlet (as a function of its mass) and the
resulting limit on the VLL 
mass is $M_{\mathrm{E}}\gtrsim 160 \mbox{ GeV}$, which represents a difference of $\sim 7\%$ in comparison to the expected limit obtained in original analysis.

\begin{table}[t]
\begin{center}
\makebox[\textwidth]{%
\begin{tabular}{|c|ccc|ccc|ccc|}
\hline
& \multicolumn{3}{c|}{$ZZ$}  & \multicolumn{3}{c|}{$WZ$} & \multicolumn{3}{c|}{$Z\gamma~ (\times 10^{-2})$ } \\ 
\multicolumn{1}{|c|}{Selection cuts}&  A  &  B   &  C    &   A  &  B     &   C   &   A   &   B    &  C     \\ \hline\hline
 \multicolumn{1}{|c|}{\begin{tabular}[c]{@{}c@{}}OSSF lepton pair with\\$|m_{\ell\ell}- m_Z| < 10\;\mathrm{GeV}$\end{tabular}}  &   0.25   &   0.25     &    0.25   &    0.18   &    0.18    &    0.18    &      0.48 &   0.48    &   0.48    \\ \hline
$\Delta R (Z,\mathrm{off-}Z\;\mathrm{lepton}) < 3$ & 0.16     & 0.16      &    0.16  &    0.10   &  0.10  & 0.10  & 0.14   & 0.14      &  0.14     \\ \hline
 $p_T^{\ell_1} > \{80,\;100,\;120\}$                     &  0.054    &   0.029   &   0.0098  & 0.029     & 0.015      &   0.0052    &    0.05   & 0.02      &     0.01   \\ \hline
 $p_T^{\ell_2} > \{20,\;40,\;60\}$                 &      0.054 &     0.025  &   0.0073  &    0.029     &   0.012  &    0.0035   &       0.05   &  0.02  &   0.01     \\ \hline
 $p_T^{\ell_3} > \{0,\;0,\;20\}$                     &     0.054  &  0.025   &   0.0067 &  0.029     &     0.012  &     0.0031  &   0.05     &   0.02   &  <0.01    \\ \hline
off-$Z$ lepton = $e$                 &     0.029    &     0.013  &  0.0034 &   0.013    &  0.0053     &    0.0014   &      0.03 &    0.01   &   <0.01    \\ \hline
$m_T < 160\;\mathrm{GeV}$ ($3ljj$)    & 0.0079     &   0.0043   &  0.0014    &  0.0013     &  0.0009     &  0.0004     &   0.01    &  0.01     &   <0.01    \\ \hline
\end{tabular}
}
\end{center}
\caption{\label{VLL:newcuts:table} Efficiencies for the background
  events after applying the selection cuts corresponding to the new
  analysis performed at $\sqrt{s} = 13 \;\mathrm{TeV}$. Efficiencies
  are presented as the number of events which are selected over the
  number of initial events. The regions represented by $\{A,B,C\}$
  correspond to different values of the mass of the VLL, $\{
  M_{\mathrm{E}} < 300,\;  300 \leq M_{\mathrm{E}} < 400,\;
  M_{\mathrm{E}} \geq 400 \}$, in GeV.
  We have included in the generation only leptonic
  decays of the $Z$ and $W$ into electron, muons and taus but apply
  our cuts only to final state electrons and muons.
} 
\end{table}

In order to see what the reach with the current recorded luminosity
can be we have repeated the same analysis at $\sqrt{s}=13\;\mathrm{TeV}$ and
$\mathcal{L} = 139 \;\mathrm{fb^{-1}}$. 
However, we can take advantage of the higher center of mass
energy to impose more stringent cuts,
in particular on the transverse momentum of the leading leptons.
Since the $p_T$
of the observed leptons in signal events increases with the increase in the VLL mass, we have defined 3
clusters of masses in which the selection threshold for $p_T$ of observed leptons
varies. We present in Table~\ref{VLL:newcuts:table} the definition of
these clusters and the efficiencies of all selection cuts.  With
  these more stringent selections, we were able to apply generation
  level cuts on the $Z\gamma$ background, generating only events in
  which at least one lepton has $p_T > 62\,\mathrm{GeV}$. 

\begin{figure}%
    \centering
{\includegraphics[width=0.48\linewidth]{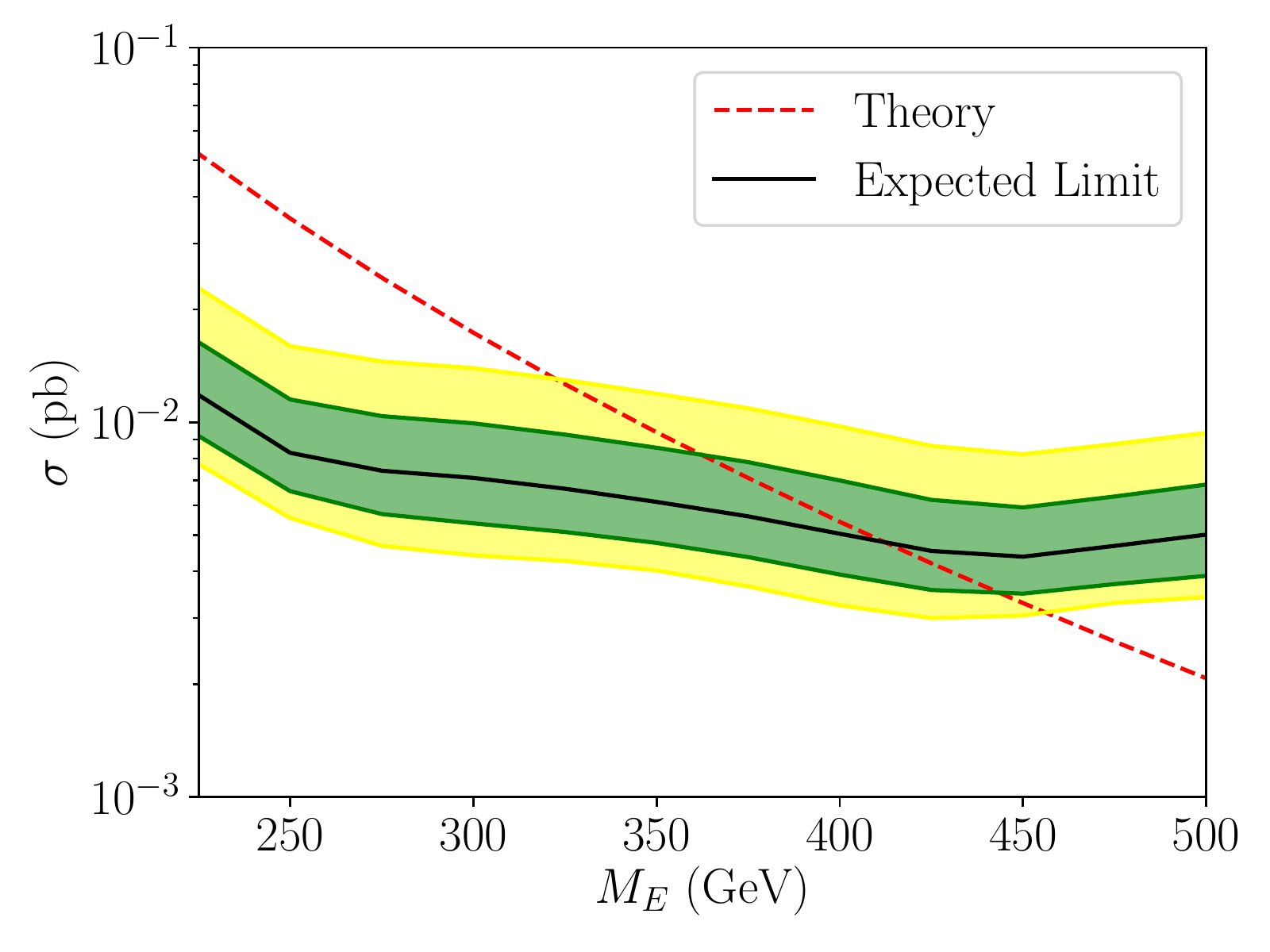} }%
    ~
{\includegraphics[width=0.48\linewidth]{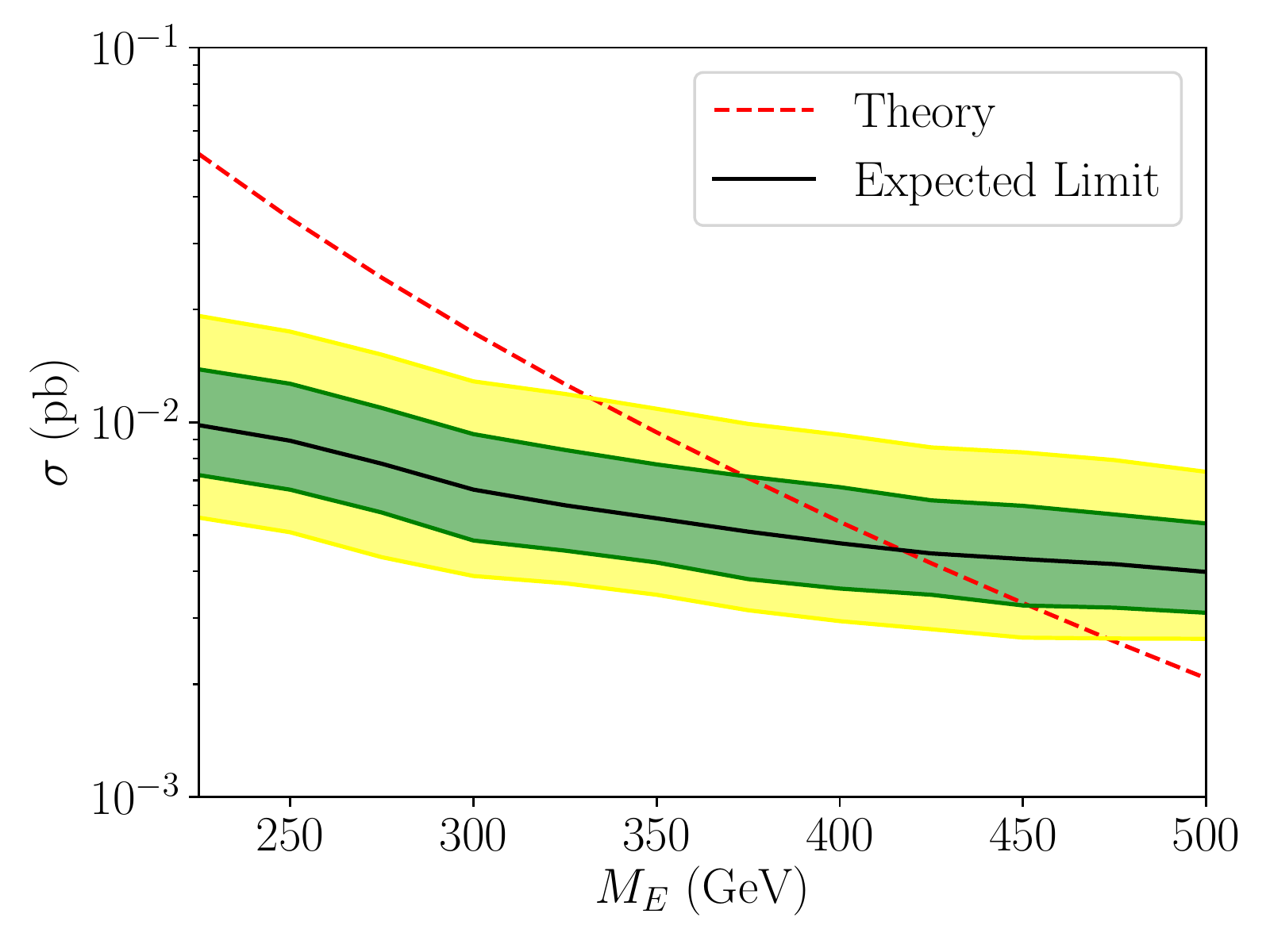} }%
    \caption{Exclusion plot for a VLL singlet decaying into electrons
      (muons) on the left (right) panel at the LHC for $\sqrt{s}=13
      \mbox{ TeV}$ and an integrated luminosity of
      $139\;\mathrm{fb}^{-1}$ using the improved analysis. See text
      for details.}%
    \label{fig:newanalysis_13tev}%
\end{figure}

Furthermore, at these higher energies, we can also remove almost the entirety of the $WZ$ background by setting a cut on the transverse mass, $m_T$, of the reconstructed $W$ boson. This cut is effective because in events from $WZ \rightarrow \ell \nu \ell \ell$, in principle, the off-$Z$ lepton is coming from the $W$ decay and the missing energy of the event, $\slashed{E}_T$, originates from the neutrino. Therefore, for events from the $WZ$ background, we have 
\begin{equation}
m_T = \sqrt{2\left(\slashed{E}_T p_{T\ell} - \mathbf{\slashed{E}_T}.\mathbf{p_{T\ell}}\right)} \leq m_W\, ,
\end{equation} 
 where $p_{T\ell}$ is the transverse momentum of the off-$Z$ lepton. As such, this quantity should, in principle, be at most the mass of the $W$ boson. In order to keep as many signal events as possible, this cut is only performed on the 3-lepton signal region, which contains most of the $WZ$ background. 
Figure \ref{fig:newanalysis_13tev} shows the limits obtained with this
improved analysis. Assuming that the observed data corresponds to the
expected background, a mass of the VLL up to $410\;\mathrm{GeV}$ ($420
\;\mathrm{GeV}$) could be excluded by this analysis for the case in
which the off-$Z$ lepton is an electron (muon). Given the
  similarity between the limits obtained when the VLL couples to first
  or second generation of SM leptons, we will only explore the case in
  which it couples to electrons hereafter.

\begin{figure}%
    \centering
    {\includegraphics[width=0.6\linewidth]{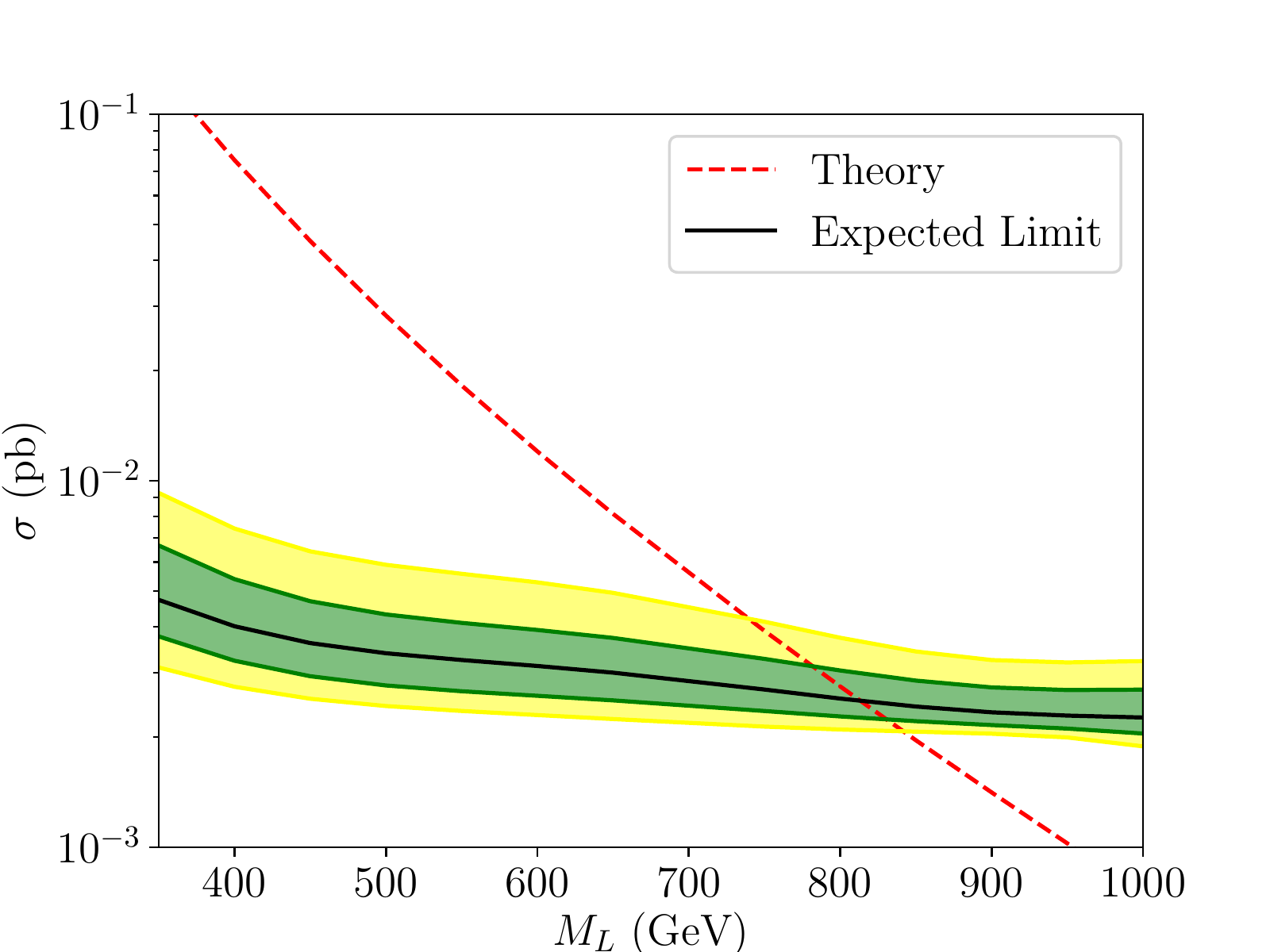} }%
    \caption{Exclusion plot for a VLL doublet decaying into electrons
      (or electron neutrinos) at the LHC for $\sqrt{s}=13 \mbox{ TeV}$
      and an integrated luminosity of $139~\mathrm{fb}^{-1}$ using our
      improved analysis.}
    \label{fig:doublet_139_77}%
\end{figure}

Despite being tailored for the case in which the VLL is a singlet of
$SU(2)$, this analysis can also be applied for a VLL doublet, $L$, of
hypercharge $-1/2$. In this case we need to take into account not only
the pair production of the charged component of the VLL doublet, $
p\;p \rightarrow E^+ E^-$, but also the pair production of the neutral
component, $p\;p \rightarrow N N$, and the associated production of
both, $p\; p \rightarrow E^\pm N$. For large masses (we will consider
both components degenerate in mass), the charged component will decay
equally to $\ell Z$ and $\ell H$, while the neutral component decays
solely to $\nu W$. Therefore, our background remains the same, and as
such we can recast the previous analysis to the doublet case,
obtaining the limits shown in Figure \ref{fig:doublet_139_77}, where we
considered the off-$Z$ lepton to be an electron. As expected, we
obtain much more stringent bounds than on the singlet case, with
masses up to $\sim720\;\mathrm{GeV}$ being excluded.

An analysis
searching for VLL doublets of hypercharge $-1/2$ was performed by the
CMS collaboration in Ref. \cite{Sirunyan:2019ofn} with an integrated
luminosity $\mathcal{L} = 77 \;\mathrm{fb}^{-1}$. A bound $M_L\geq
790\;\mathrm{GeV}$ was obtained in this analysis due to a statistical
fluctuation in the observed data. The expected limit in that analysis,
which is the fair comparison to the bound we can compute,
corresponded to $M_L^{\mathrm{expected}}\geq 690\;\mathrm{GeV}$. Rescaling our search to
the same integrated luminosity we find
$M_L^{\mathcal{L}=77\;\mathrm{fb}^{-1}}\geq 730\;\mathrm{GeV}$,
remarkably close to the expected limit in the CMS
search, despite the fact that the CMS analysis targets decays into 
tau leptons and therefore a direct comparison is not straight-forward.

\subsubsection{Decays with missing energy}

\begin{figure}[h!]
\centering
  \includegraphics[width=0.5\linewidth]{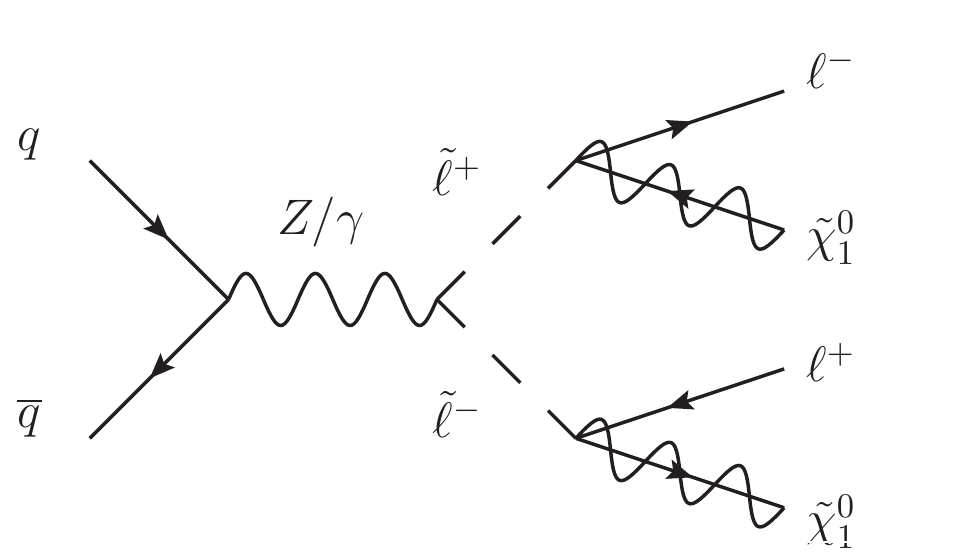}
  \caption{Pair production and decay of sleptons at hadron
    colliders. $\tilde{\ell}$ represents a charged slepton and
    $\tilde{\chi}^0_1$ the lightest neutralino. }
  \label{fig:slep_process}
\end{figure}

To explore the case in which the VLL decays predominantly into
a SM lepton and missing energy ($A_H$ in our case), we consider an ATLAS
analysis~\cite{Aaboud:2018jiw} at $\sqrt{s} = 13 \; \mathrm{TeV}$ and
an integrated luminosity of $\mathcal{L} = 36.1\; \mathrm{fb}^{-1}$
searching for pair produced sleptons decaying into a SM lepton and a
neutralino, as represented in Figure \ref{fig:slep_process}. The
analysis selects events with 2 OSSF leptons ($e$, $\mu$), imposing a
veto on additional jets. It also rejects events with an invariant mass
of the two leptons $m_{\ell\ell} < 40 \;\mathrm {GeV}$. Several
inclusive and exclusive signal regions are defined in which different
requirements are demanded for $m_{\ell\ell}$ and the $m_{T2}$
variable~\cite{Barr:2003rg,Lester:1999tx}, defined by 
\begin{equation}
m_{T2} =
\mathrm{min_{\mathbf{q_{T}}}}[\mathrm{max}(m_T(\mathbf{p_{T1},q_T}),
  m_T(\mathbf{p_{T2},\slashed{E}_T
    - q_T})]\, , 
\label{eq:stransverse}
\end{equation}
where $\mathbf{p_{T1,2}}$ represent the transverse momentum of each of
the identified leptons and $\mathbf{q_T}$ is the vector that minimizes
the maximum of both transverse masses, defined as 
\begin{equation}
m_T(\mathbf{p_T,q_T})  = \sqrt{2(p_Tq_T - \mathbf{p_Tq_T})} \, .
\end{equation}

\begin{figure}[h!]
\centering
  \includegraphics[width=0.6\linewidth]{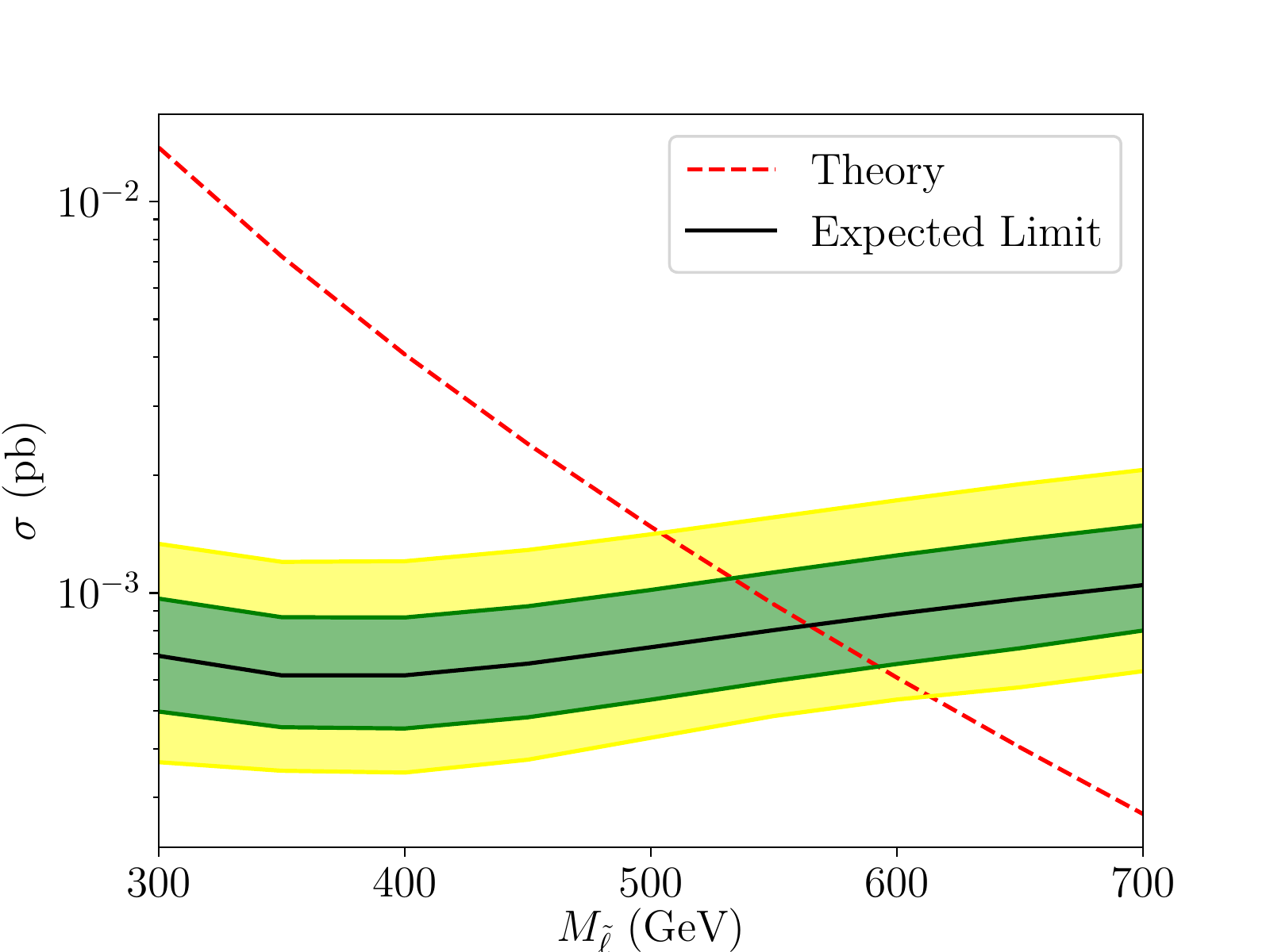}
  \caption{Limit on the slepton mass, $M_{\tilde{\ell}}$, for a
    neutralino mass $M_{\tilde{\chi}^0_1}=1\;\mathrm{GeV}$ following
    the analysis in~\cite{Aaboud:2018jiw}.}
  \label{fig:limits_sleps}
\end{figure}

The main backgrounds for this signal are diboson processes ($ZZ$,
$WW$ and $WZ$) and $t\bar{t}$. In order to maximize our statistics
we always include 2 leptons (with an $m_{\ell\ell} > 95\;\mathrm{GeV}$
partonic cut) in the final state in the background
generation. These 2 leptons can be of either family for the diboson
case and are restricted to the first two generations for the
$t\overline{t}$ one.
The selection cuts and corresponding efficiencies are presented in
Table~\ref{slepton:cuts:table:13}. We have validated the 
analysis including all signal regions proposed in the original
  analysis~\cite{Aaboud:2018jiw}. However, new results are calculated
  considering only the signal region of $m_{\ell\ell} >
  111\;\mathrm{GeV}$ and $m_{T2} > 100\;\mathrm{GeV}$ as we find the
  difference in regards to all signal regions not
  significant. 
Given that the analysis in Ref.~\cite{Aaboud:2018jiw} applies to
sleptons and neutralinos, which are scalars and fermions,
respectively, as opposed to our case with a VLL and a dark photon
(fermion and vector, respectively), in order to validate our analysis
we have implemented a slepton-neutralino model. Fixing the neutralino
mass $M_{\tilde{\chi}^0_1}=1\;\mathrm{GeV}$ we obtain the results
shown in Figure~\ref{fig:limits_sleps}, with a limit
$M_{\tilde{\ell}}\geq 565\; \mathrm{GeV}$, very similar to the expected
limit obtained by the ATLAS collaboration, $\sim 570\;\mathrm{GeV}$. 

\begin{table}
\begin{center}
 \begin{tabular}{|c c c c c |} 
 \hline
Selection Cuts & $ZZ $ & $WZ $ & $WW$ &$t\overline{t}$\\ [0.5ex]  
\hline\hline
\multicolumn{1}{|c}{\begin{tabular}[c]{@{}c@{}}OSSF lepton pair\end{tabular}}& 0.33 & 0.33 & 0.23 &  0.53\\ 
 \hline
$m_{\ell\ell} > 40\;\mathrm{GeV}$ & 0.31 & 0.28 & 0.11 &0.53\\
 \hline
 $p^{\ell_1}_T > 25\;\mathrm{GeV}$ & 0.31 & 0.28 & 0.11& 0.53\\
\hline
 $p^{\ell_2}_T > 20\;\mathrm{GeV}$ & 0.28 & 0.26 & 0.095&0.51\\
 \hline
 $p^{b-jet}_T < 20\; \mathrm{GeV}$ & 0.24 & 0.24 & 0.093 & 0.13\\
\hline
 $p^{jet}_T < 60\; \mathrm{GeV}$ & 0.15 & 0.14 & 0.081 &0.061\\
 \hline
$m_{T2} > 100 \; \mathrm{GeV}$ &0.0064 &0.003 &0.0002 & 0.0001\\
\hline
$m_{\ell\ell} > 110 \; \mathrm{GeV}$ & 0.0016 &0.001 & 0.0002&0.0001\\
\hline
\end{tabular}
\caption{\label{slepton:cuts:table:13} Cumulative efficiencies for the
  background events after applying the selection cuts corresponding to
  the analysis performed at $\sqrt{s} = 13 \;\mathrm{TeV}$ for slepton searches
  ~\cite{Aaboud:2018jiw}. Efficiencies are presented as the number of
  events which are selected over the number of initial events (see
  text for details).}
\end{center}
\end{table}

\begin{figure}[h!]
\centering
  \includegraphics[width=0.6\linewidth]{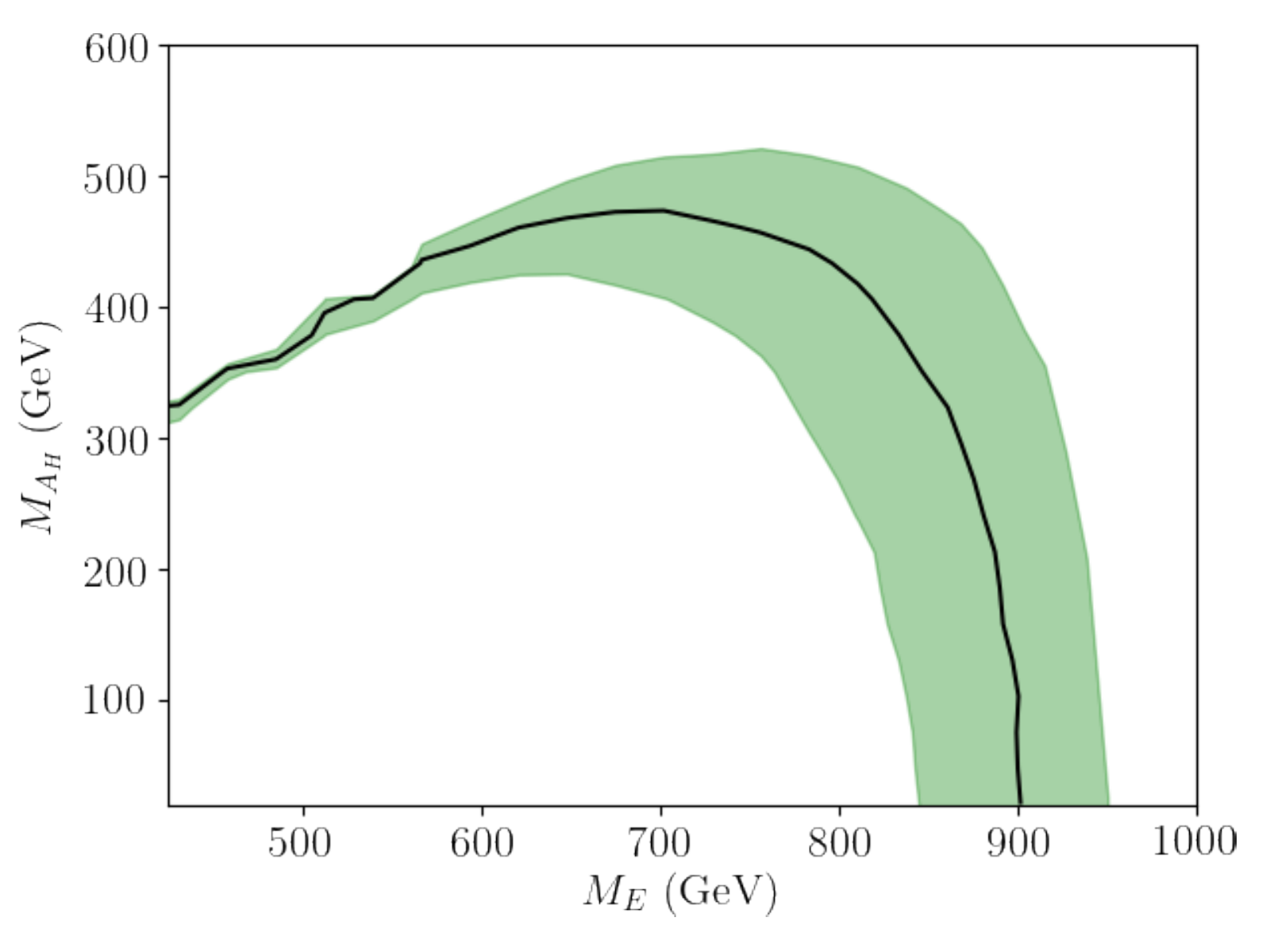}
  \caption{Expected (solid black) and 1-sigma band (green)
    $95\%$ C.L. exclusion limit in the $M_E-M_{A_H}$ plane from the
    analysis in~\cite{Aaboud:2018jiw}. The excluded region is the one
  below the curves.}
  \label{fig:2dplot_sleps}
\end{figure}

Once we have validated the analysis we can apply it to the VLL model.
Contrary to the case of purely SM decays, in this case we have a 
new degree of freedom in our analysis, the mass of the other new
particle, $M_{A_{H}}$. Some models predict this mass to be close to
the electroweak scale, such as the Littlest Higgs model with T-parity,
but in other cases, we can have sub-GeV masses as
is the case in feebly interacting massive particles (FIMP) in which
this new particle plays the role of DM as we will see
below. As such for each mass point of the VLL,  we vary $M_{A_{H}}$
from 1 GeV up to the mass of the VLL in question. We present the
limits obtained for $\sqrt{s}=13\;\mathrm{TeV}$ and an integrated
luminosity $\mathcal{L} = 139 \; \mathrm{fb}^{-1}$ in
Figure~\ref{fig:2dplot_sleps}.  
As expected, the analysis is more constraining for lighter $A_H$. As
the mass difference between $A_H$ and the VLL decreases, the leptons
from signal events become softer and more difficult to identify and
pass the selection cuts. For $M_{E} \gtrsim 900\; \mathrm{GeV}$ the
production cross-section is too low and the analysis cannot constrain
the signal regardless of $M_{A_{H}}$.

\subsection{Constraints on vector-like leptons with general decays}
Once we have ensured the accuracy of our simulations to recast the
experimental searches in the limiting cases in which the VLL decays
only through SM or missing energy channels, we are in a position to
interpolate between them and therefore consider the case of arbitrary
branching fractions in the different channels. To do this, we have
scanned different possible branching ratios for each of the decay
channels. In order to not need to generate every signal corresponding
to different BRs, we apply a weight to each signal event according to
its decay. To do this, we generate a signal of Drell-Yan pair-produced
VLLs with $\mathrm{BR}_g(E \rightarrow A_H \ell ) =  \mathrm{BR}_g(E \rightarrow Z \ell
)= \mathrm{BR}_g(E \rightarrow H \ell ) = \mathrm{BR}_g(E \rightarrow W\nu) = 0.25$,
where the $g$ subscript describes the generated sample. To probe a
specific point with different BRs, each event is weighted by $\mathrm{BR}_p^i /
\mathrm{BR}_g^i$, where $p$ subscript represents the probed branching ratio and
$i$ superscript corresponds the specific decay -- the decay of each
event is determined at generator level.  

\begin{figure}%
    \centering
    {\includegraphics[width=0.45\linewidth]{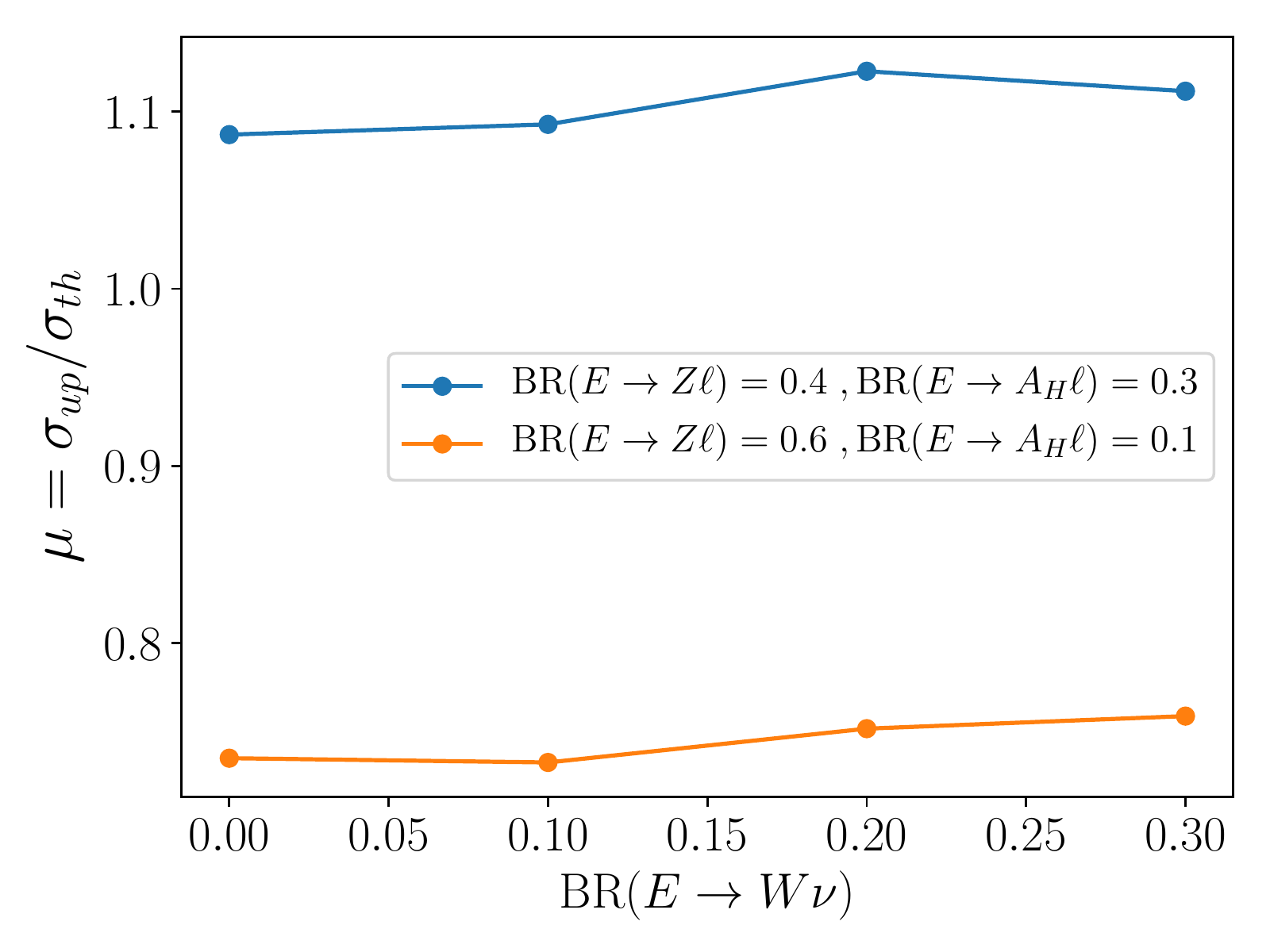} }%
    \qquad
    {\includegraphics[width=0.45\linewidth]{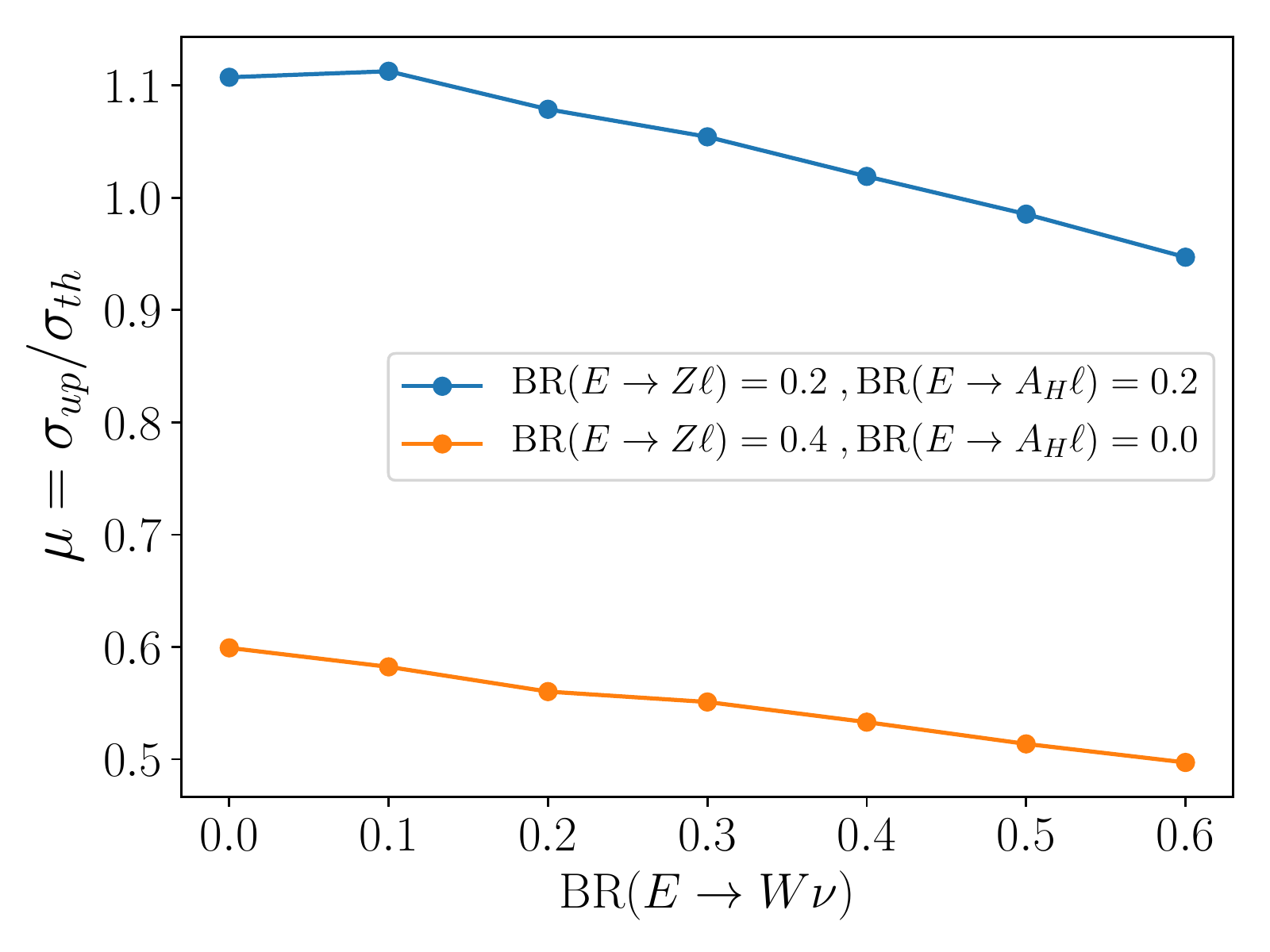} }%
    \caption{Dependence of the signal strength, $\mu$, on $\mathrm{BR}(E\to W \nu)$ for different values of the remaining parameters. The masses are taken to be $M_E = 500 \; \mathrm{GeV}$ and $M_{A_H} = 98\;\mathrm{GeV}$ ($M_E = 400 \; \mathrm{GeV}$ and $M_{A_H} = 98\;\mathrm{GeV}$) for the left (right) panels. Values of $\mu$ smaller than 1 are considered excluded. }%
    \label{fig:dependence_Wnu}%
\end{figure}

Which analysis is more constraining depends on the particular value of the branching ratios but since they specifically target the final states with either $\ell A_H$ and $\ell Z$, the results are presented in the $\mathrm{BR}(E \rightarrow A_H \ell ) \;\mathrm{vs}\; \mathrm{BR}(E \rightarrow Z \ell )$ plane with the others branching ratios being fixed to 
\begin{equation}
\mathrm{BR}_p(E\rightarrow W\nu) = 2 \mathrm{BR}_p(E\rightarrow H\ell) = \frac{2}{3} \left[ 1 - \mathrm{BR}_p(E\rightarrow A_H \ell) - \mathrm{BR}_p(E\rightarrow Z\ell) \right],
\end{equation}
which correspond to the relation between the different branching
ratios in the large $M_E$ limit for $\mathrm{BR}(E\to A_H
\ell)=0$.
We have checked that the corresponding bounds are quite insensitive to this latter choice. The residual dependence is due to cross-contamination between different channels into our signal regions. However, this effect is small as shown in Figure~\ref{fig:dependence_Wnu}, where we represent the change in the signal strength $\mu$ as a function of the branching ratio into $W\nu$ for two different values of the remaining parameters. The signal strength that represent our discriminating variable changes by $20\%$ at most, which results in a very mild dependence of the final limit on $M_E$.

\begin{figure}%
    \centering
    \includegraphics[width=0.45\linewidth]{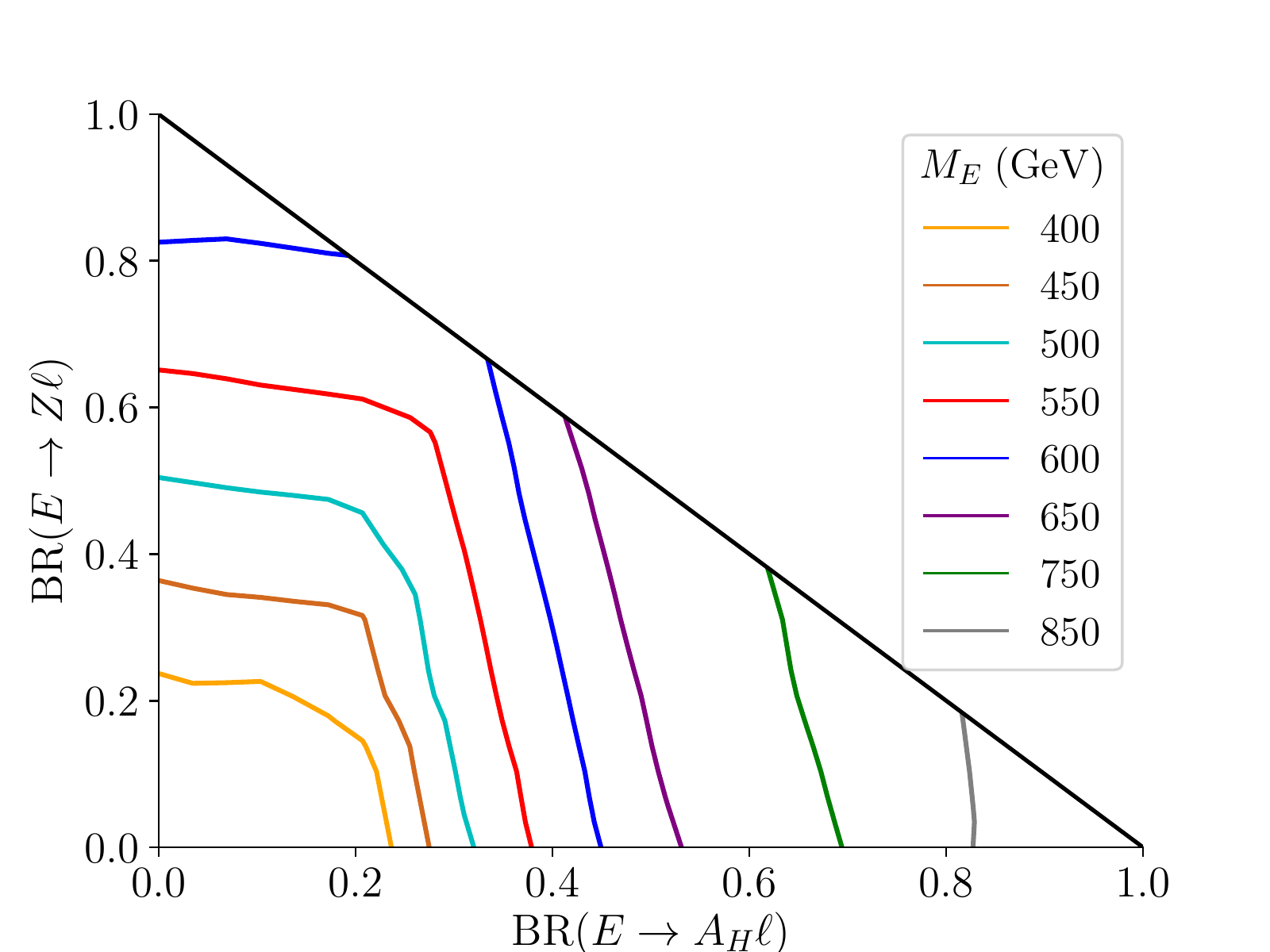}%
    \qquad
   \includegraphics[width=0.45\linewidth]{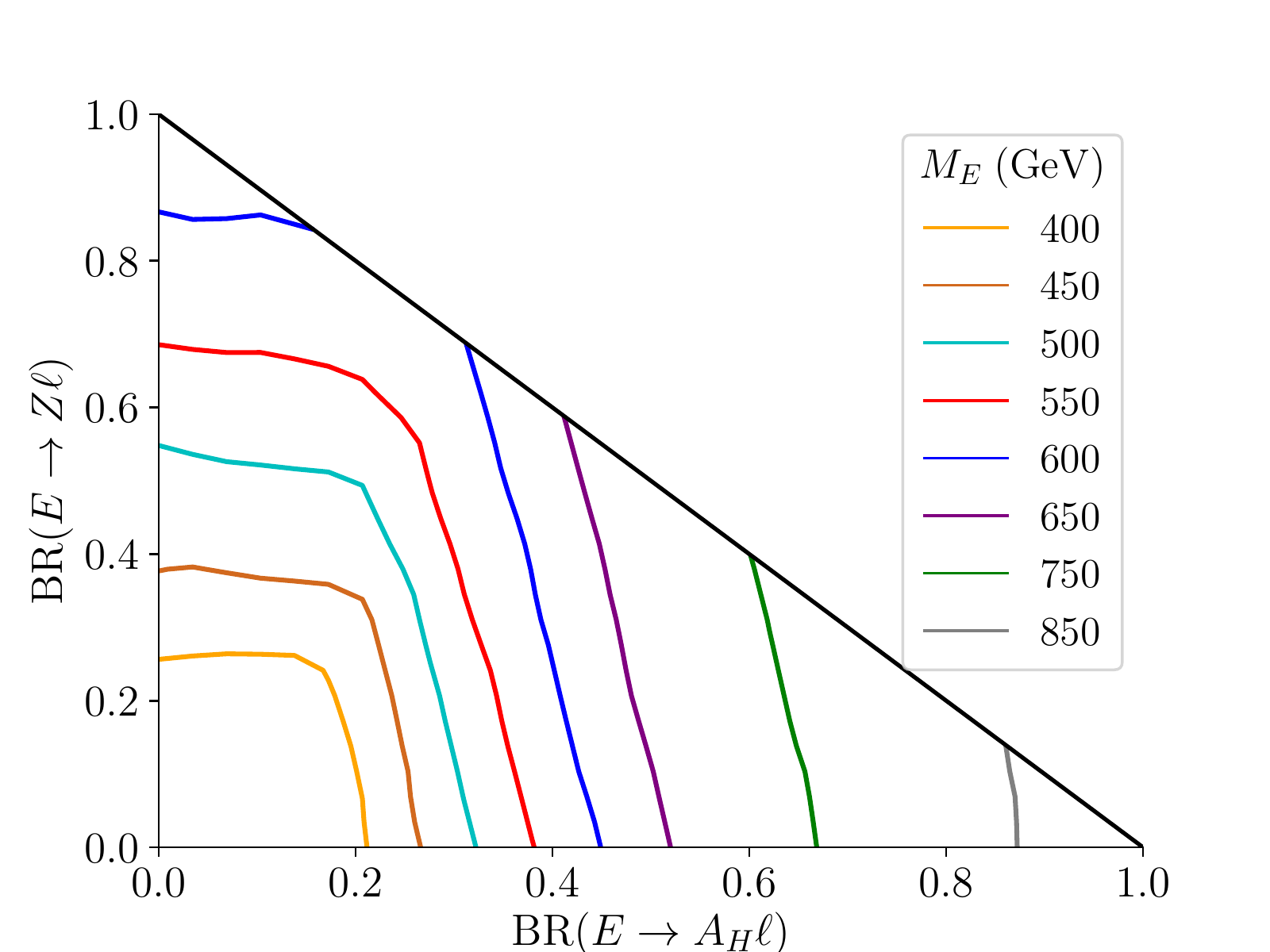}%
    \caption{$95\%$ C.L. lower bound on the VLL mass $M_E$ as a
      function of $\mathrm{BR}(E\to A_H \ell)$ and $\mathrm{BR}(E\to Z \ell)$ for
      $M_{A_H}=1\;\mathrm{GeV}$ (left panel) and
      $M_{A_H}=98\;\mathrm{GeV}$ (right panel).
      The limits are given as
      contour plots for fixed values of $M_E$ in which the region
      above and to the right of the curves is excluded and they are
      computed from a
      combination of the analyses described in the previous section
      with $\sqrt{s}=13\,\mathrm{TeV}$ and an integrated luminosity of
      $\mathcal{L}=139\;\mathrm{fb}^{-1}$. The contours correspond to
      masses that grow from the $(0,0)$ vertex outwards.}%
    \label{fig:triangle_139}%
\end{figure}

Our final result, that combines the two analyses discussed in the
previous section for arbitrary values of $\mathrm{BR}(E\to A_H \ell)$ and
$\mathrm{BR}(E\to Z \ell)$ are shown in Figure~\ref{fig:triangle_139} for two different
values of $M_{A_H}=1\;\mathrm{GeV}$ (left panel) and
$M_{A_H}=98\;\mathrm{GeV}$ (right panel). As expected, the effect of
the $A_H$ mass is more relevant in the region in which the missing
energy signal dominates and for lighter values of the VLL mass, since
the smaller mass difference results in a softer lepton. Still, except
for very low branching ratios into the decay channels targeted by our
analysis, the differences are minimal. Thus, from now on we will only
report our results for $M_{A_H}=1\;\mathrm{GeV}$.  
We show the results as contours for fixed value of $M_E$ with the
region above and to the right of each contour line being excluded for
that mass at the $95\%$ CL. The limit for the VLL singlet case with SM
decays can be easily obtained by considering the vertical axis, which
corresponds to $\mathrm{BR}(E\to A_H \ell)=0$, at the relevant (mass dependent)
$\mathrm{BR}(E\to Z \ell)$. The most
stringent bounds are along both axes, when the branching ratios into the
channels we are most sensitive to are maximized. The numerical value
of the limits in these three interesting cases are 
\begin{equation}
 M_{E}\gtrsim \left \{ \begin{array}{l}
                         405\;\mathrm{GeV}, \quad [\mbox{VLL singlet}], \\
                         630\;\mathrm{GeV}, \quad [\mathrm{BR}(E\to \ell Z)=1], \\
                         895\;\mathrm{GeV},\quad [\mathrm{BR}(E\to \ell A_H)=1],
                        \end{array}
                        \right. \quad [\sqrt{s}=13\;\mathrm{TeV}, \mathcal{L}=139\;\mathrm{fb}^{-1}].
\label{eq:139results}
\end{equation}

\begin{figure}%
    \centering
    \includegraphics[width=0.75\linewidth]{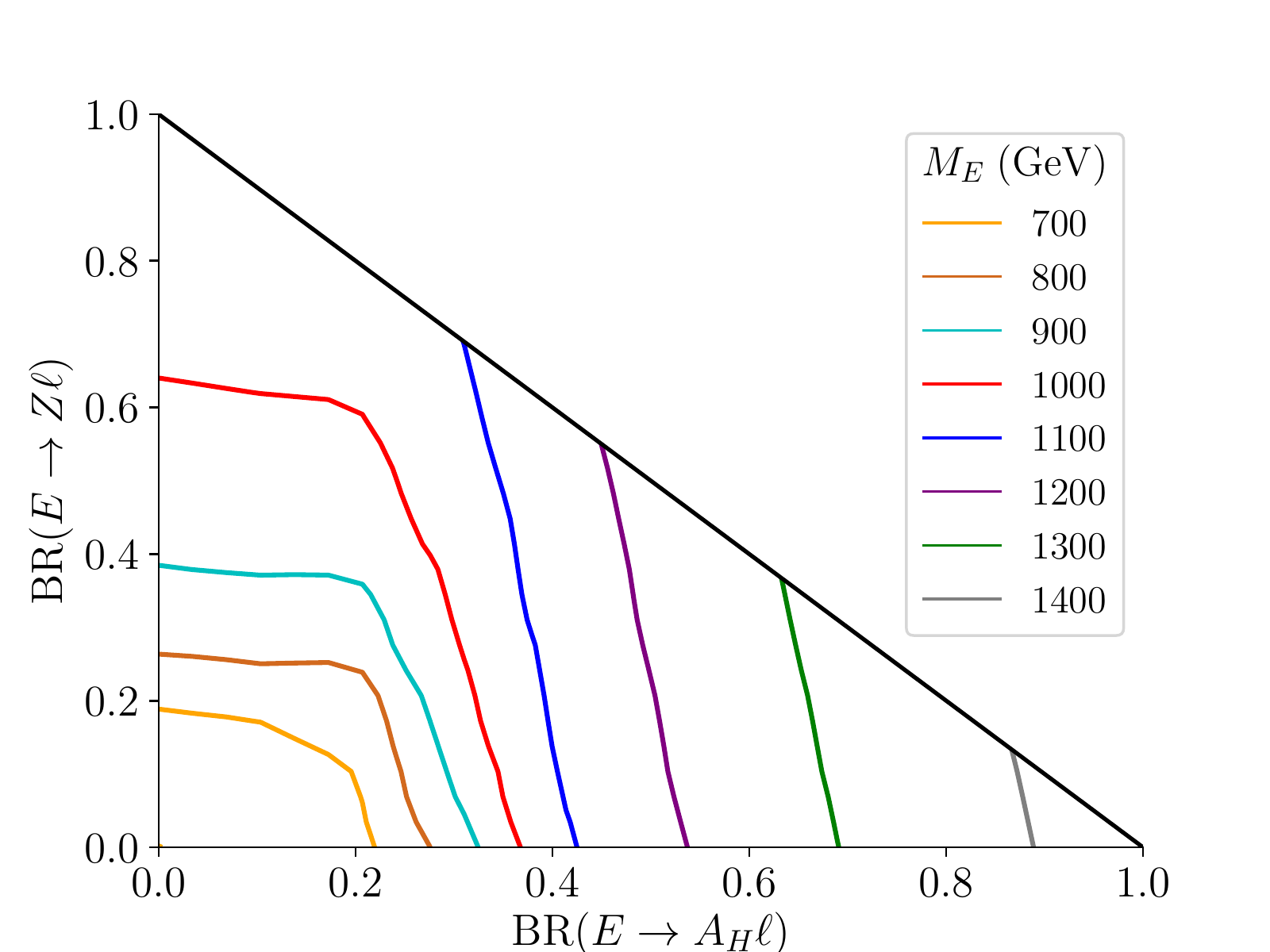} %
    \caption{Projected limits on the mass of the VLL, $M_E$ for
      arbitrary branching fractions for the HL-LHC. (See text and
      Figure~\ref{fig:triangle_139} for details.)}
    \label{fig:triangle_3ab}%
\end{figure}

\subsection{Future projections}

The constraints presented in Figure~\ref{fig:triangle_139} represent
the current constraints on a new charged VLL with general decays. In
this section we explore the potential of the LHC to probe new VLLs in
its high-luminosity (HL-LHC) and high-energy (HE-LHC)
configurations. We will also explore the potential reach of the 100
TeV hh-FCC.  

Starting with the HL-LHC (for which we take
$\sqrt{s}=13\;\mathrm{TeV}$ and an integrated luminosity of
$\mathcal{L}=3\;\mathrm{ab}^{-1}$) we use the same improved analysis
described in the previous section and in
Table~\ref{VLL:newcuts:table}, making sure 
that we generate enough statistics for the required integrated
luminosity. The result is shown, for $M_{A_H}=1\;\mathrm{GeV}$,
in Figure~\ref{fig:triangle_3ab}. The
correspond final reach of the HL-LHC in the limiting cases of a VLL
singlet, $\mathrm{BR}(E \to \ell Z)=1$ and
$\mathrm{BR}(E\to \ell A_H)=1$ is, respectively,
\begin{equation}
 M_{E}\gtrsim \left \{ \begin{array}{l}
                         785\;\mathrm{GeV}, \quad [\mbox{VLL singlet}], \\
                         1090\;\mathrm{GeV}, \quad [\mathrm{BR}(E\to \ell Z)=1], \\
                         1450\;\mathrm{GeV},\quad [\mathrm{BR}(E\to \ell A_H)=1],
                        \end{array}
                        \right. \quad [HL-LHC].
\end{equation}

\begin{figure}%
    \centering
    \includegraphics[width=0.75\linewidth]{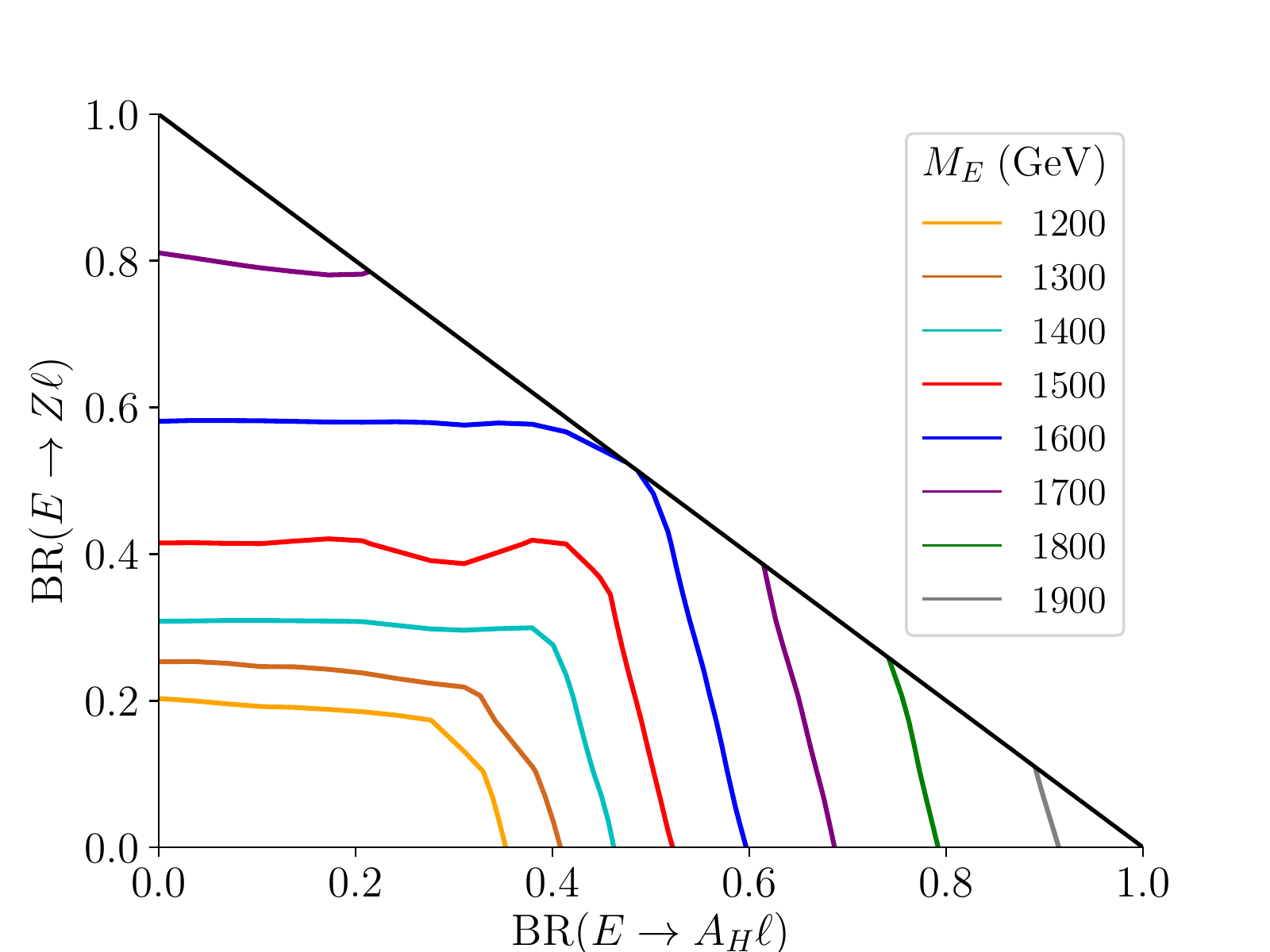} %
    \caption{Projected limits on the mass of the VLL, $M_E$ for
      arbitrary branching fractions for the HE-LHC. (See text and
      Figure~\ref{fig:triangle_139} for details.)}
    \label{fig:triangle_HE-LHC}%
\end{figure}

When considering a higher energy collider, like the HE-LHC, for which
we consider $\sqrt{s}=27\;\mathrm{TeV}$ and
$\mathcal{L}=3\;\mathrm{ab}^{-1}$, we can again afford to
impose more stringent cuts on the different variables involved in the
analysis, in particular in the lepton $p_T$. In the SM decays
analysis, we impose a partonic cut on all backgrounds of $p_T > 75
\;\mathrm{GeV}$ of the leading lepton whereas for the analysis
focusing on the missing energy decay, backgrounds were generated with
a partonic cut of $p_T > 100 \;\mathrm{GeV}$ for the leading
lepton. We were able to use this cut since we updated the selection
thresholds from Table ~\ref{slepton:cuts:table:13} to $p_T^{\ell_1} >
120\;\mathrm{GeV}$ in the missing decay analysis. The resulting reach,
again for $M_{A_H}=1\;\mathrm{GeV}$,
is reported for arbitrary branching ratios in
Figure~\ref{fig:triangle_HE-LHC}. The estimated reach, in the limiting
cases is 
\begin{equation}
 M_{E}\gtrsim \left \{ \begin{array}{l}
                         1295\;\mathrm{GeV}, \quad [\mbox{VLL singlet}], \\
                         1770\;\mathrm{GeV}, \quad [\mathrm{BR}(E\to \ell Z)=1], \\
                         1965\;\mathrm{GeV},\quad [\mathrm{BR}(E\to \ell A_H)=1],
                        \end{array}
                        \right. \quad [\text{HE-LHC}].
\end{equation}

\begin{figure}%
    \centering
    \includegraphics[width=0.75\linewidth]{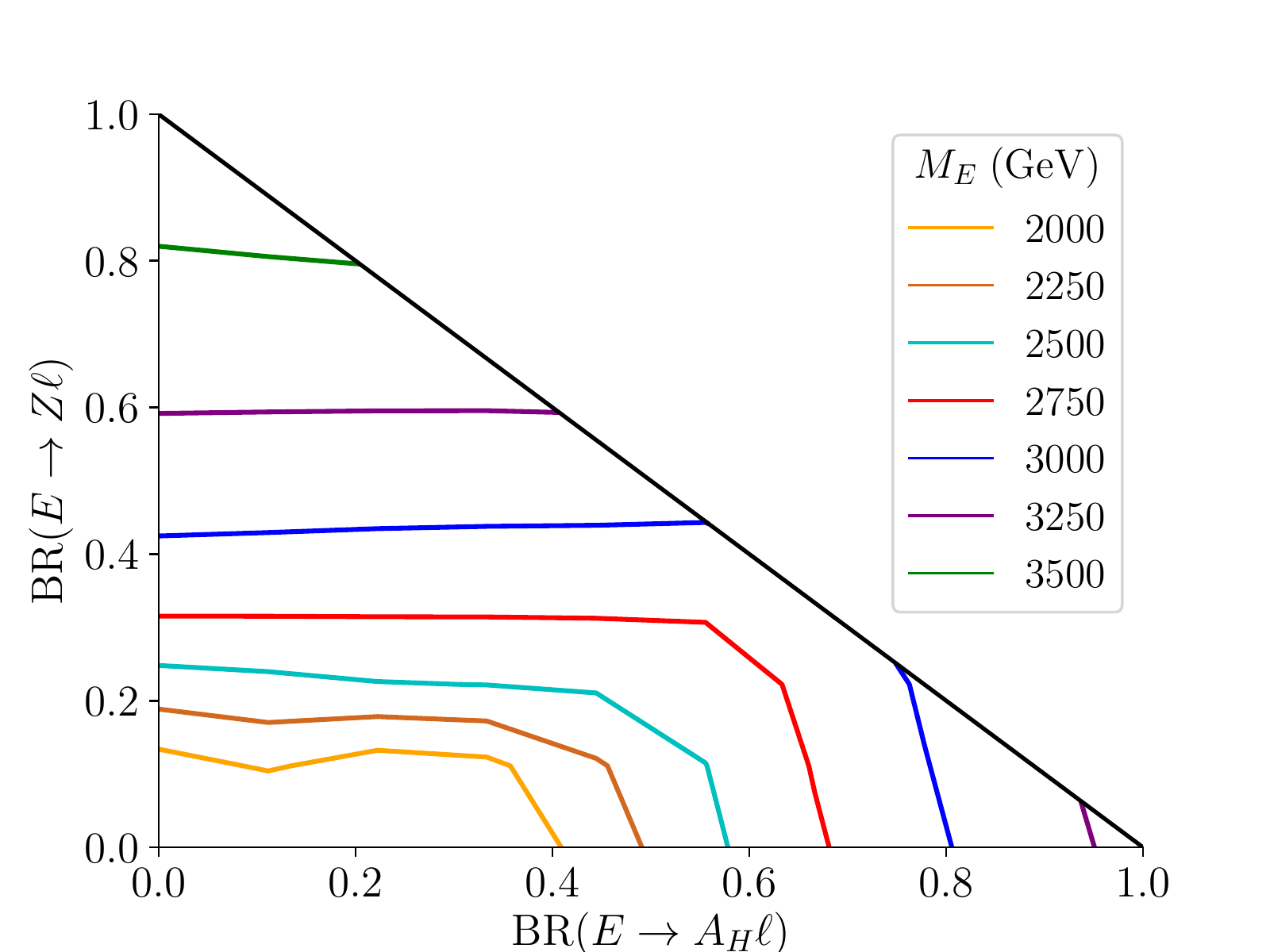} %
    \caption{Projected limits on the mass of the VLL, $M_E$ for
      arbitrary branching fractions for the 100 TeV hh-FCC. (See text and
      Figure~\ref{fig:triangle_139} for details.)}
    \label{fig:triangle_FCC}%
\end{figure}

A detailed study of the reach of future circular colliders on VLLs
with general decays is beyond the scope of the present work, however,
we can use a crude estimate of the corresponding reach at the hh-FCC
by considering the instantaneous luminosity as used in the
\texttt{Collider Reach} tool~\cite{colliderreach}. First we
test the validity of this approach by extrapolating the current
luminosity results reported in Eq. (\ref{eq:139results}) to the HL-LHC
and to the HE-LHC. We find that the extrapolation agrees with our
detailed simulation within $6\%(14\%)$  in the case of the HL-LHC for
the SM decays analysis (missing decays analysis) and within
$6\%(35\%)$ for the HE-LHC for the SM decays analysis (missing decays
analysis). The latter case shows the differences that arise not only
from the increased production cross sections of signal and backgrounds
but also from the more stringent cuts that we can imposed with higher
energy. The difference in the missing decays analysis drops to 14 \%
when we extrapolate from the HL-LHC results. We can expect a similar
effect when extrapolating our results to the FCC. Assuming
$\sqrt{s}=100\;\mathrm{TeV}$ and $\mathcal{L}=3\;\mathrm{ab}^{-1}$ we
obtain the results shown in Figure~\ref{fig:triangle_FCC} and the
following limits in the VLL, pure $Z$ and pure $A_H$ decay cases

\begin{equation}
 M_{E}\gtrsim \left \{ \begin{array}{l}
                         2525\;\mathrm{GeV}, \quad [\mbox{VLL singlet}], \\
                         3665\;\mathrm{GeV}, \quad [\mathrm{BR}(E\to \ell Z)=1], \\
                         3330\;\mathrm{GeV},\quad [\mathrm{BR}(E\to \ell A_H)=1],
                        \end{array}
                        \right. \quad [\text{hh-FCC (extrapolation)}].
\end{equation}

\begin{figure}%
    \centering
    \includegraphics[width=0.75\linewidth]{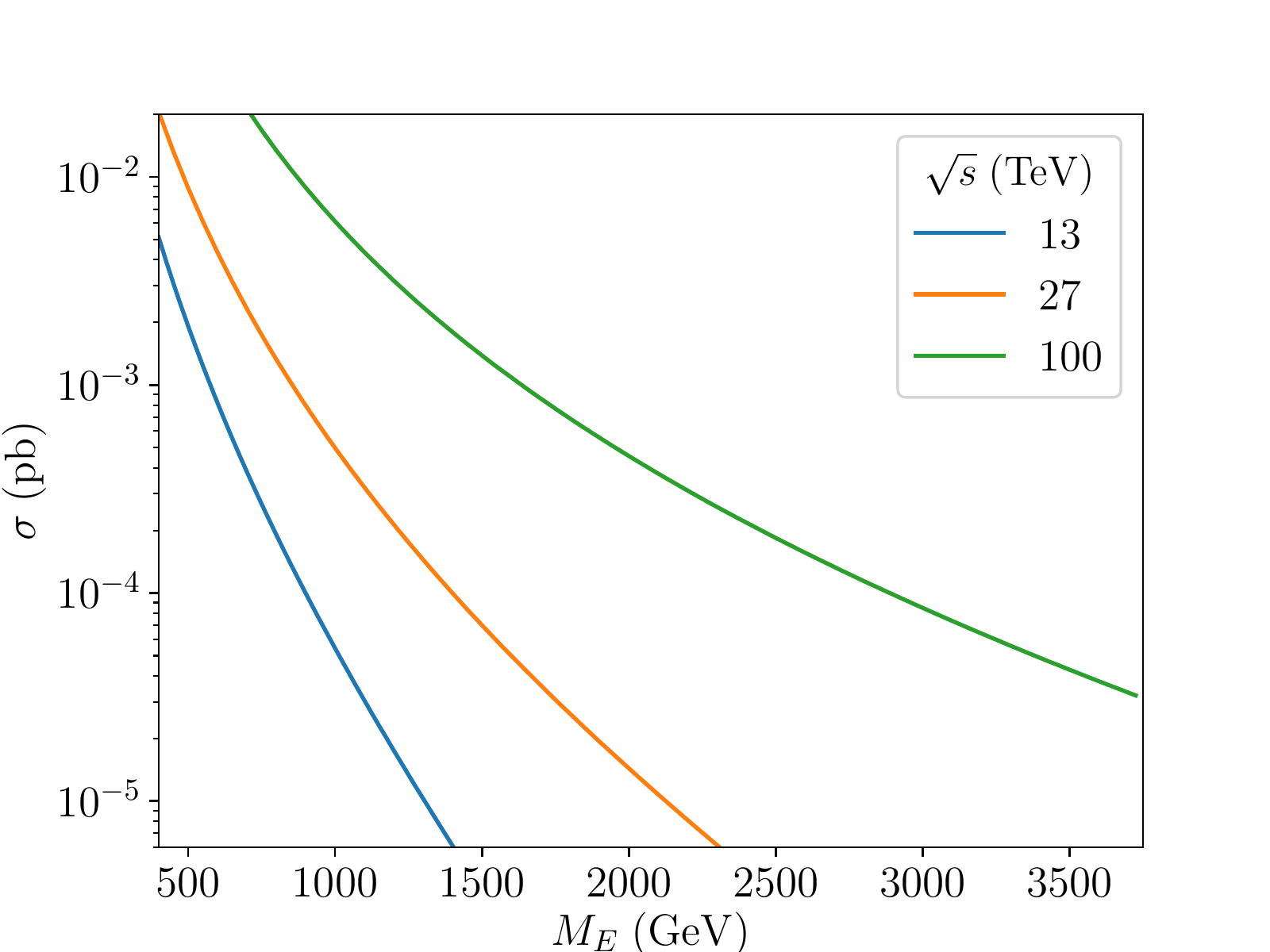} %
    \caption{VLL singlet pair production cross-section at ($pp$) hadron colliders with $\sqrt{s} = 13\,,27\,,100\,\mathrm{TeV}$.}
    \label{fig:cross_sections}%
\end{figure}

The results reported in Figures~\ref{fig:triangle_139}-\ref{fig:triangle_FCC} are completely general except for the fact that we are using the production cross-section of a VLL singlet with hypercharge -1 to obtain the mass limits. For the sake of generality, we provide in Figure~\ref{fig:cross_sections} the cross-sections we have used for the LHC, HE-LHC and hh-FCC so that our limits can be applied to more general VLLs by rescaling the corresponding pair production cross-section.

\section{Dark photon as a dark matter candidate\label{sec:dm}}

So far we have just assumed that the lifetime of $A_H$ is large enough
to appear as missing energy at detector scales. However, if $A_H$ has
a lifetime larger than the age of the universe, it becomes a suitable
candidate for DM. As such, we can use the observed relic density and
direct detection experiments to further constrain these models. In
this section we will focus on two possible production mechanisms for
DM. We will first consider the case in which $A_H$ has a mass around
the electroweak 
scale and its abundance is fixed through the freeze-out
mechanism. Then we will consider the possibility that $A_H$ is light
and has a very weak coupling to the SM so that its production follows
the freeze-in mechanism.

\subsection{Standard freeze-out}

For the case of a heavy DM candidate -- with a mass around the
eletroweak scale -- we will consider that it is stabilized through a
symmetry. An example of this arises in the Littlest Higgs model with
T-parity (LHT)~\cite{Cheng:2004yc,Low:2004xc},
in which $A_H$ is T-odd, as is the vector-like lepton,
while the SM particles are T-even. Therefore, the VLL decays
exclusively through the missing energy channel.
Since $A_H$ is a singlet of the SM, we can write the following
operators 
\begin{align}
\label{eq:lagdm}
 \mathcal{L}&= - c_{A_Hh} g^{\prime\,2} \left( \sqrt{2}v h A_H^\mu A_{H_{\mu}} +
 \frac{1}{2} h h A_H^\mu A_{H_{\mu}} \right) + q_H g^\prime [\bar{E}_L
   \gamma_\mu \ell_R + \mathrm{h.c.}] A_H^\mu+\ldots\,, 
\end{align}
where $v\approx 174\;\mathrm{GeV}$, the dots represent other couplings that are irrelevant for the
viability of $A_H$ as a DM candidate and we have included explicit factors of
the $U(1)_Y$ gauge coupling $g^\prime$ to make the connection with the
LHT model more direct. In the LHT model $c_{A_Hh} = \frac{1}{8} $ and
$q_H =\frac{1}{10}$ \cite{delAguila:2008zu}.

\begin{figure}[h!]
\centering
  \includegraphics[width=\linewidth]{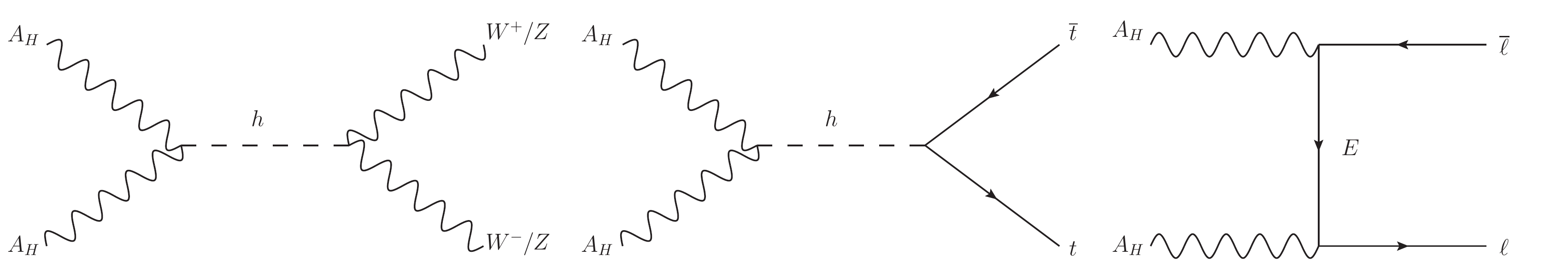}
  \caption{Relevant annihilation processes for the dark photon in the
    standard freeze-out mechanism.}
  \label{fig:anprocess}
\end{figure}

The latest Planck results measured the relic density abundance to be
$\Omega h^2 \sim 0.12$ \cite{Aghanim:2018eyx} 
 and therefore the model must predict a relic density equal to ($A_H$
 accounts for all of DM) or smaller than ($A_H$ is only part of DM
 content) that number. The most relevant processes for the
 annihilation of $A_H$ 
 are to b-quarks, $W^+ W^-$ or $Z$ bosons or top quarks (depending on
 the mass of the DM candidate) through the s-channel exchange of a
 Higgs \cite{Birkedal:2006fz}. Furthermore the annihilation into leptons through the exchange
 of the VLL is also important -- the corresponding diagrams are shown in
 Figure~\ref{fig:anprocess}. 
 Therefore, as mentioned above, the relic density calculation will be
 controlled by the 
 couplings of $A_H$ to the Higgs and the coupling to the VLL and SM
 lepton and thus we will scan different values for these
 couplings. Given that the VLL mediates one of these channels, when
 the s-channel annihilation is subdominant, the mass difference
 between $A_H$ and the VLL will also play an important role.

\begin{figure}
\centering
  \includegraphics[width=0.9\linewidth]{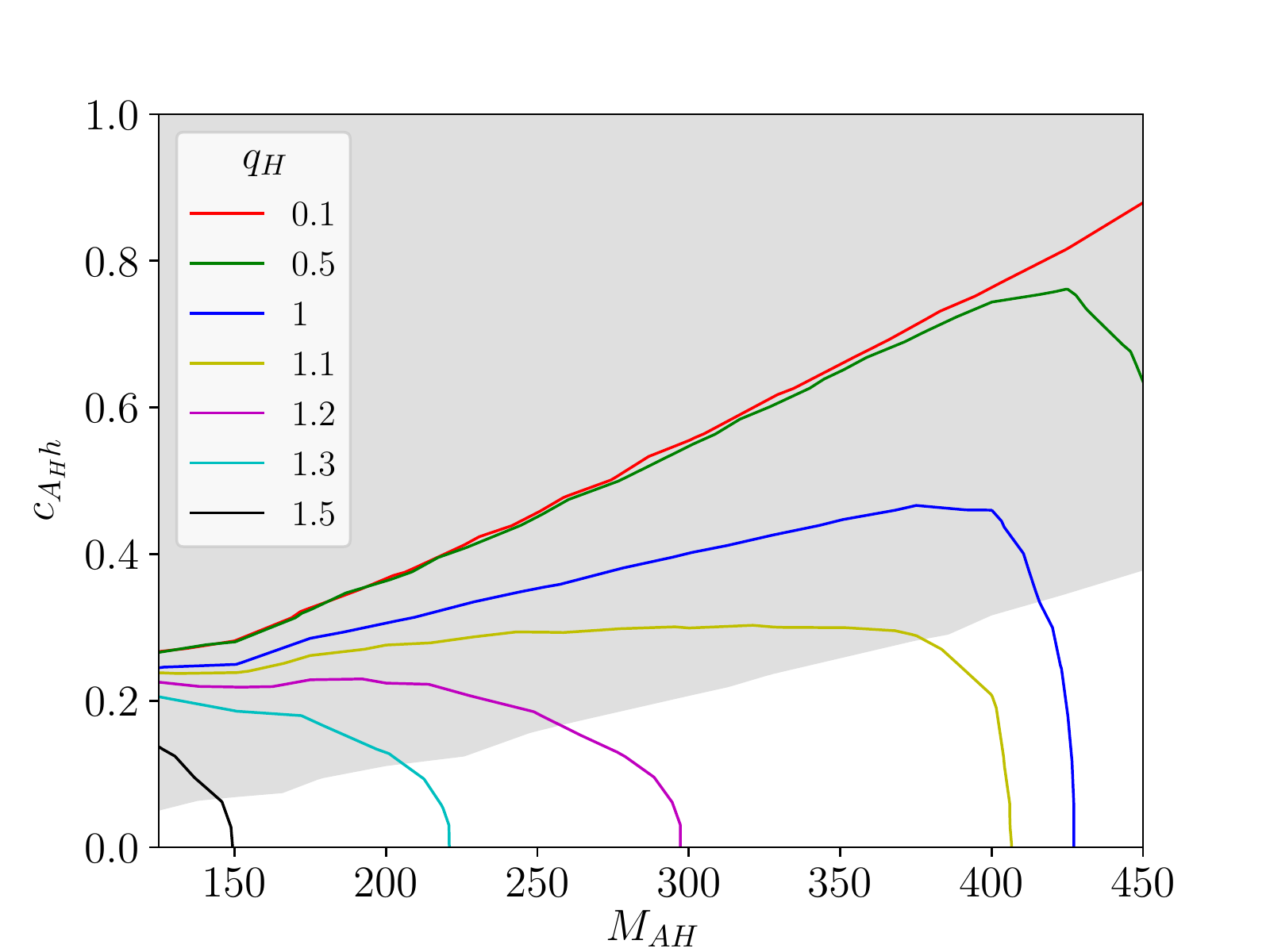}
  \caption{Contours of $\Omega h^2=0.12$ for fixed values of
    $q_H$. The region below the curves gives too large DM density and
    is excluded. The shaded region is excluded from direct detection
    experiments. See text for details.}
  \label{fig:annresults}
\end{figure}

\begin{figure}[h!]
\centering
  \includegraphics[width=0.35\linewidth]{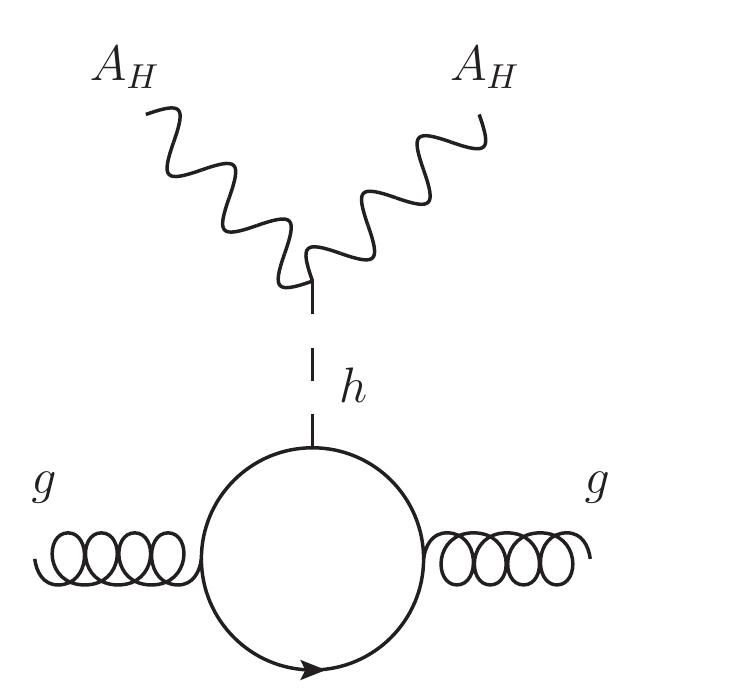}
  \caption{Relevant direct detection process.}
  \label{fig:ddprocess}
\end{figure}

The calculation of the relic density is done using
\texttt{MadDM}~\cite{Ambrogi:2018jqj} by 
inputting a \texttt{UFO} model~\cite{Degrande:2011ua} which we generate through
\texttt{Feynrules}~\cite{Alloul:2013bka}.
The results are presented in Figure~\ref{fig:annresults} for a VLL mass 
$M_E=500\;\mathrm{GeV}$ in the $M_{A_H}-c_{A_Hh}$ plane for different
values of $q_H$. The curves represent the values for which the relic
abundance agrees with the observed value. The region below the curve
is excluded and the one above requires further sources of DM. We also
show, shaded in grey, the excluded region from direct detection using
the latest XENON1T data~\cite{Aprile:2018dbl}.  
For small values of $q_H$ ($<1$), the
s-channel annihilation through the Higgs dominates; however, as we
increase $q_H$, the channel mediated by the VLL becomes more important
and we get a significant rise in the annihilation cross-section, with
a $q_H \sim 1.5$ allowing almost all of the depicted parameter
space. As expected, we can also see (particularly for high enough
values of
$q_H$) that, as the mass difference between the VLL and the DM
candidate decreases, the impact of the VLL-mediated channel increases.  
The coupling to the Higgs boson is also important
for the spin-independent scattering cross-section with nucleons, as
the dominant diagrams are the Higgs exchange with quarks or with
gluons through a loop of heavy quarks as represented in
Figure~\ref{fig:ddprocess}. We have computed the corresponding
scattering cross-section with \texttt{MadDM} and shown the excluded
region in Figure~\ref{fig:annresults}. 

Varying $q_H$ also affects direct detection constraints. In principle
$q_H$ could be responsible for a 1-loop DM nucleon scattering
amplitude, mediated by a photon. However, as noted in
Ref. \cite{Agrawal:2014ufa}, for the 
case of a real DM vector candidate, the coupling between 2 DM
particles and a photon will be described by a dimension-6 operator,
since the dimension-4 $A_{H_{\mu}} A_{H_{\nu}} F^{\mu\nu}$ does not
exist due to the antisymmetry of the field strength tensor. Moreover,
the resulting amplitude will be further suppressed when one takes the
non-relativistic limit. As such, we will neglect contributions from
this process to direct detection bounds in this work. 

Another experimental observable which may be affected by changing $q_H$ is
the anomalous magnetic moment of both the electron and the muon,
depending on which of these SM leptons the VLL couples to. The latest
experimental results are ~\cite{Aoyama:2012wj,Muong-2:2021ojo} 
\begin{align}
\Delta a_e &= a_e^{\mathrm{Exp}} - a_e^{SM} = -1.06 (0.82)\times 10^{-12} \; , \\
\Delta a_\mu & = a_\mu^{\mathrm{Exp}} - a_\mu^{SM} = 25.1 (5.9) \times 10^{-10} \; ,
\end{align}
where the uncertainties include theoretical and experimental contributions.

\begin{figure}[h!]
\centering
  \includegraphics[width=0.7\linewidth]{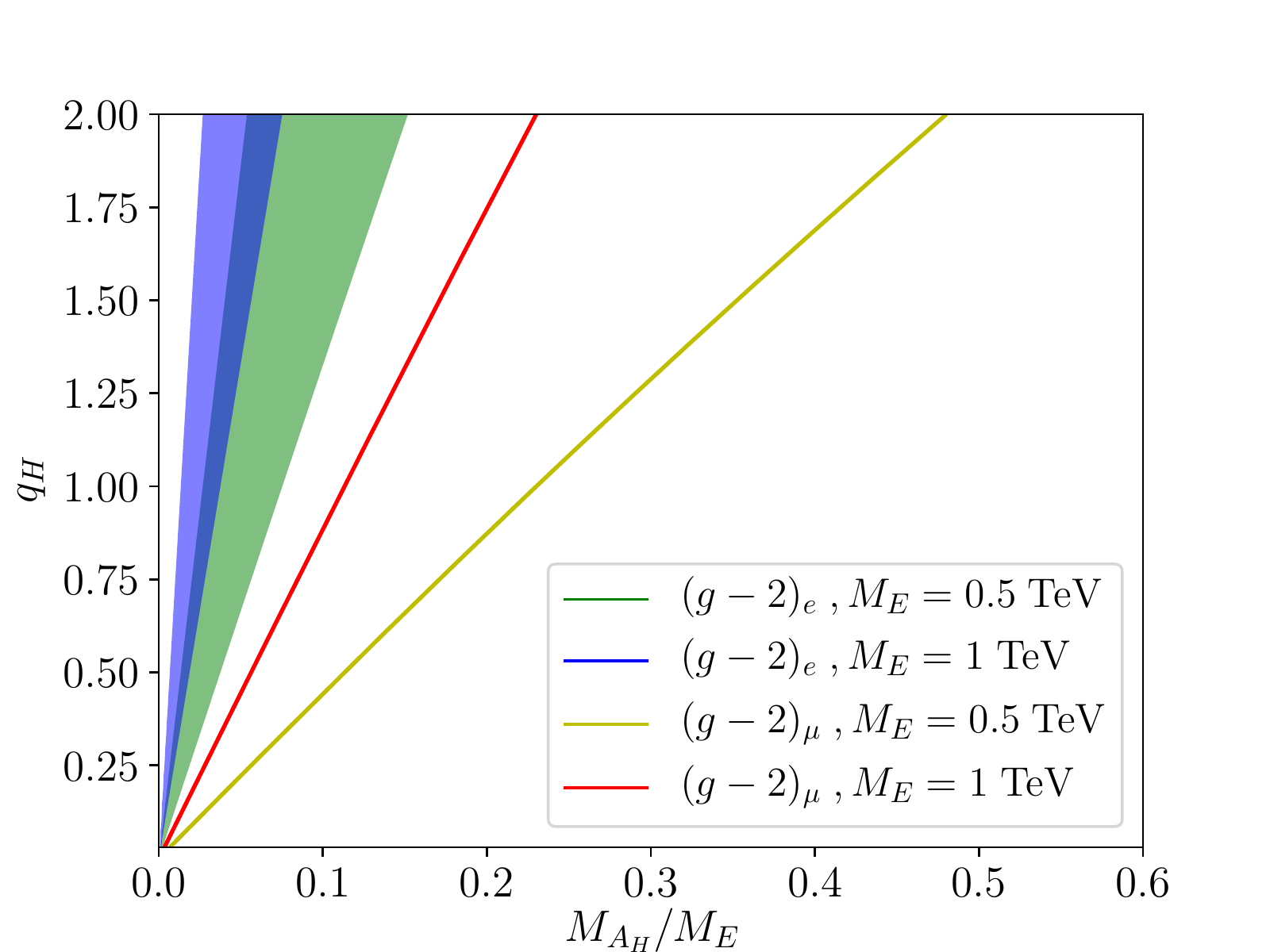}
  \caption{Region that explains the electron $g-2$ anomaly for $M_E=500\; \mathrm{GeV}$ (green) and $M_E=1\,\mathrm{TeV}$
    (blue). The 2 sigma limits from the contribution to the
    muon $g-2$ are shown with solid lines, for $M_E=500\; \mathrm{GeV}$ (yellow) and for $M_E=1\,\mathrm{TeV}$ (red). The region above the lines is excluded.}
  \label{fig:g2}
\end{figure}

The new contribution from $E$ and $A_H$ reads~\cite{Agrawal:2014ufa},
\begin{equation}
a_\ell = - \frac{\epsilon^2}{48\pi^2r^2(1-r^2)^4}q_H^2 g^{\prime\,2} \left[5-14r^2+39r^4-38r^6+8r^8+18r^4 \mathrm{ln}(r^2) \right]  + \mathcal{O}(\epsilon^3) \;,
\end{equation}
where $\epsilon \equiv m_\ell / M_{E}$ and $r\equiv M_{A_{H}} /M_{E}$
and $m_\ell$ is the mass of the SM lepton for which the contribution
is being calculated. This result is always negative and as such, it
contributes in the direction of explaining the $(g-2)$ anomaly of the
electron, whereas it goes in the wrong direction for the muon anomaly.
Figure~\ref{fig:g2} shows the parameter space that is constrained by
these measurements. For the case in which the VLL couples to
electrons, we show the region which explains the observed anomalous
magnetic moment. For the muon case, as this model increases the
tension with the experimental result we constrain this contribution to
be smaller than the combination of the experimental and theoretical
uncertainties. The region above the curves is excluded for the muon case.

The results shown in Figure~\ref{fig:annresults} reflect the well known
tension between the production of the correct relic abundance and
direct detection experiments for a standard weekly interacting massive
particle. Such tension can be relaxed
if the masses of the new particles are nearly degenerate (with the VLL being
slightly heavier). This regime of co-annihilation~\cite{Baker:2015qna}
increases the annihilation cross-section since 
processes such as $A_H E\rightarrow SM\;SM$ and $A_H SM\rightarrow
E\;SM$ can now contribute significantly. The importance of these
contributions will be a function not only of this degeneracy in mass,
but also of the coupling $q_H$. An estimation of the needed mass
splitting to have a significant 
contribution to the annihilation process can be obtained by
considering that, at the freeze-out temperature, $T_F$,  both
particles are still in equilibrium. For co-annihiliation to be
important, one would have $M_{A_{H}} - M_{E} \sim T_F$. Knowing that
$M_{A_{H}} \sim 25\;T_F$, for cold DM, the splitting must be at most
$\Delta \sim 0.04$ where $\Delta \equiv (M_{E} - M_{A_{H}})/M_{A_{H}}$. 

\begin{figure}[h!]
\centering
  \includegraphics[width=0.9\linewidth]{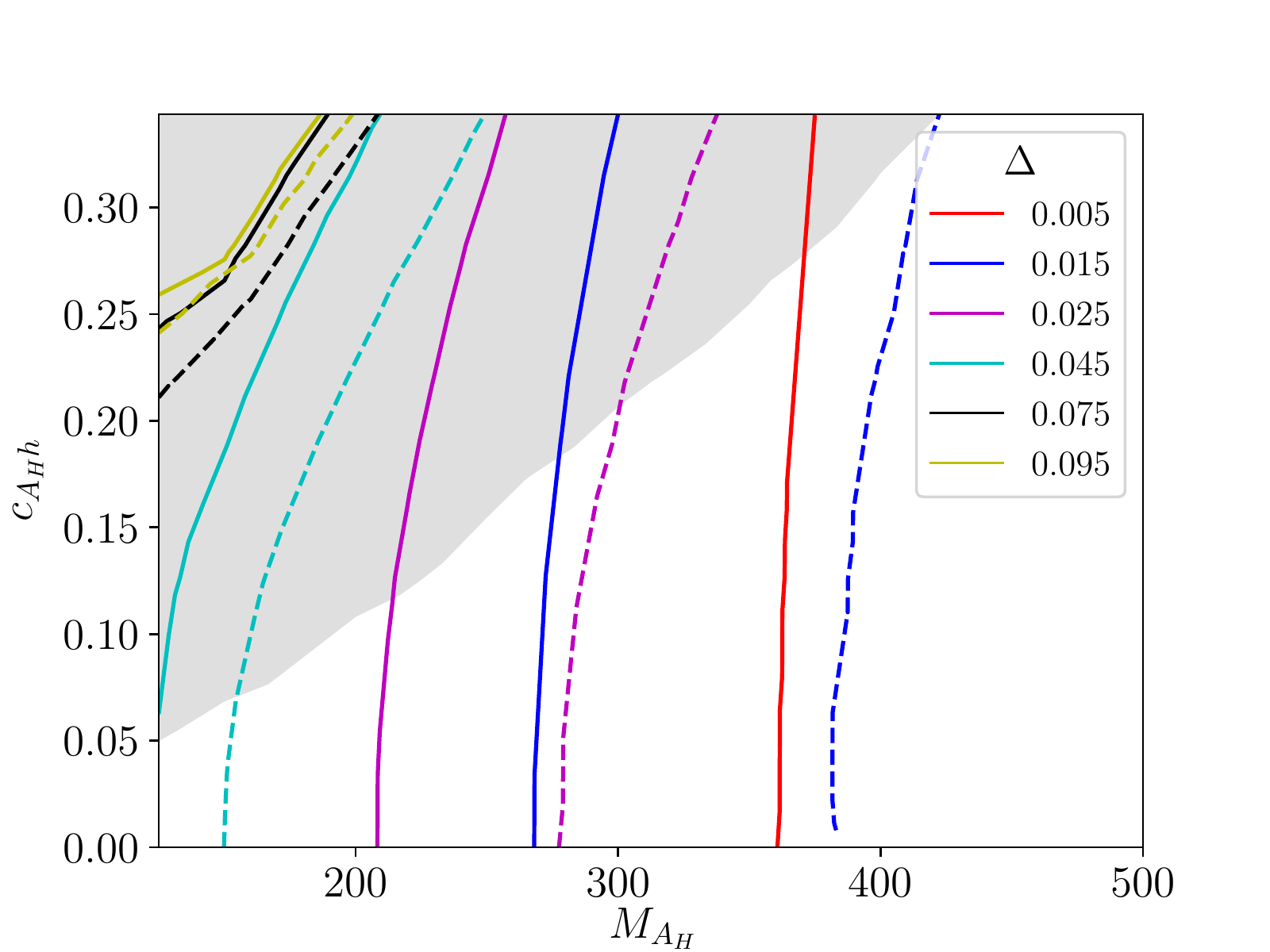}
  \caption{Contours of $\Omega h^2=0.12$ for fixed values of
    $\Delta$. The region to the right of the curve gives too large DM
    density and 
    is excluded. The solid (dashed) curves correspond to $q_H=0.1$
    ($q_H=0.2$). The shaded region is excluded from direct detection
    experiments. See text for details.}
  \label{fig:coann}
\end{figure}

In Figure~\ref{fig:coann} we show the relic density abundance for
cases in which co-annihilation can be important. We consider two
values of $q_H=0.1$ (solid) and $q_H=0.2$ (dashed) and plot the
contours of $\Omega h^2=0.12$
for different values of $\Delta$. The region to the right of
the different curves is excluded (as it gives too large relic
abundance). Again we show in shaded grey the region excluded by direct
detection experiments. We see that only
for $\Delta \lesssim 0.05$ a significant difference with respect to the
standard annihilation scenario is observed.

\begin{figure}[h!]
\centering
  \includegraphics[width=0.9\linewidth]{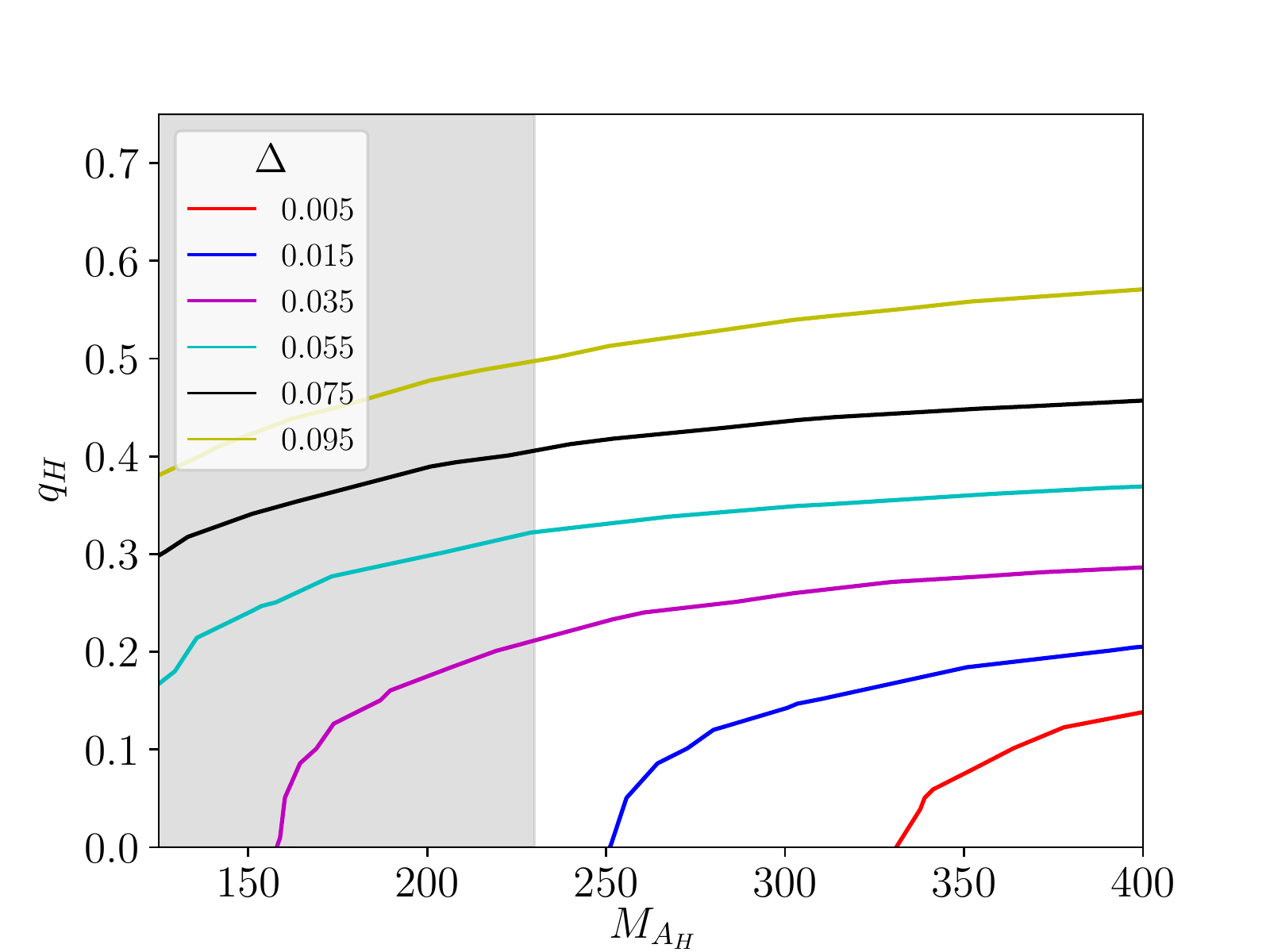}
  \caption{Contours of $\Omega h^2=0.12$ for fixed values of
    $\Delta$ for $c_{A_H h}=1$. The region below each curve gives too large DM
    density and is excluded.
    The shaded region is excluded from direct detection
    experiments. See text for details.}
  \label{fig:coann_qh}
\end{figure}

In order to better understand the dependence on $q_H$ in this
co-annihilation regime we show, in Figure~\ref{fig:coann_qh},
the $\Omega h^2=0.12$
contours in the $M_{A_H}-q_H$ plane, again for different values of
$\Delta$ for $c_{A_Hh}=1$. The region excluded by direct detection
experiments is, as usual, shaded in grey.
While for fixed $q_H$ we observed that increasing
$\Delta$ collapses the relic density line into the non co-annihilation
regime, this does not happen in this plot. In this case, even though
co-annihilation effects can be negligible for $\Delta \gtrsim 0.05$, the
annihilation process mediated by the heavy lepton is important for low
mass differences between the VLL and $A_H$ and large $q_H$ and as such
the annihilation cross-section is influenced by changes in $\Delta$
even outside the co-annihilation regime.

Note that this co-annihilation case is complementary to what was
studied in the previous section at colliders. In this case, given the
small mass difference, the final state leptons at colliders will be
very soft and therefore are very difficult to identify. There is an
ongoing effort to search for this cases of compressed mass states at
colliders ~\cite{ATLAS:2019lng} namely in the context of sleptons. Our
results show that the interpretation of such a search in the context
of VLLs is very well motivated.

\subsection{Freeze-in in feebly interacting dark matter}

In the case that DM is very light and couples very weakly to
other particles, its relic density can be set by the freeze-in
mechanism~\cite{Hall:2009bx}. In this case, the DM candidate is not in
equilibrium 
with the thermal bath but is actually produced through the decay of
other heavy particles, in our case, the decay of the VLL. This
possibility has been recently explored in~\cite{Delaunay:2020vdb} with
emphasis on the DM phenomenology, thus setting the VLL mass to a
conservative $M_E=1\;\mathrm{TeV}$ in order to avoid any collider
constraint.
In this subsection we aim to show the complementarity
between DM experiments and the collider results we presented
before for a FIMP. 

This scenario is realized by the explicit model that we present in
Appendix~\ref{realization}, to which we refer the reader for the details.
The relic density can be calculated as ~\cite{Delaunay:2020vdb}:
\begin{equation}
\Omega h^2 \approx 0.12\times 10^{-9} \frac{M_E}{M_{A_H}} \left(\frac{g_H s} {5.3\times 10^{-17}}\right) ^2,
\label{eq:relicdensity_fimp}
\end{equation}
where $g_H$ and $s$ are defined in Eqs. (\ref{eq:gh}) and (\ref{eq:s}), respectively.

For each value of $M_E$ and $M_{A_H}$ the condition that $A_H$
corresponds to all of DM, \textit{i.e.}
Eq. (\ref{eq:relicdensity_fimp})
$=0.12$, fixes the product $g_H s$. Choosing then a value for $s$ fixes all
branching ratios of the VLL. As such, by choosing a particular $s$ we
can get the collider bound from our previous analysis. 

\begin{figure}[h!]
\centering
  \includegraphics[width=0.9\linewidth]{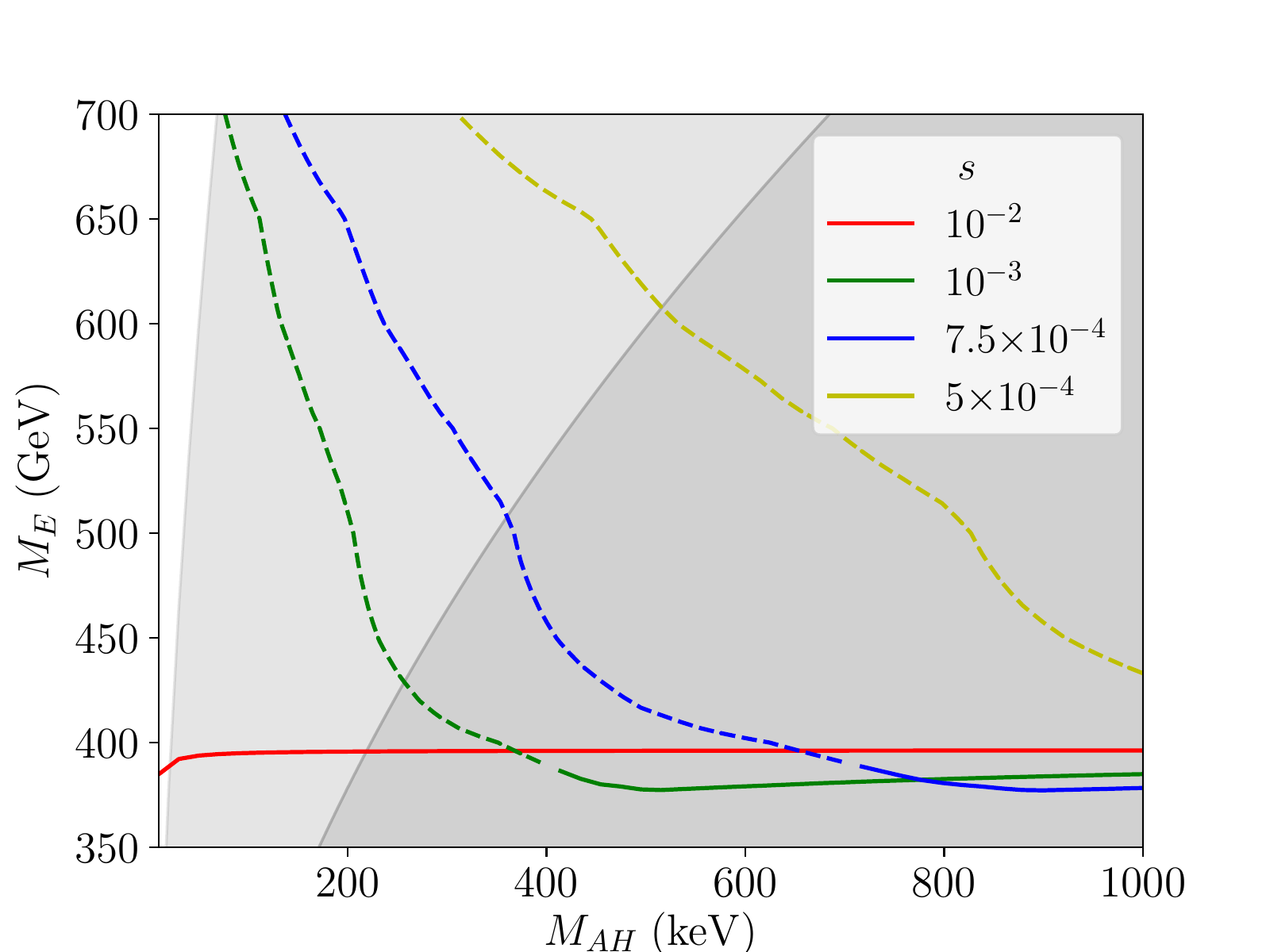}
  \caption{Contours of collider limits, for the LHC with
    $\sqrt{s}=13\;\mathrm{TeV}$ 
    and $\mathcal{L}=139\;\mathrm{fb}^{-1}$, on models that generate the
    observed DM relic abundance via the freeze-in mechanism for
    different values of the fermion mixing parameter $s$. The
    region below the curves is excluded by collider searches, either
    targeting SM decays (solid) or decays into missing energy
    (dashed). The light (dark) shaded region correspond to an
    $E\rightarrow A_H \ell$ decay length $\geq 1\;\mathrm{cm}$ ($\geq
    5\;\mathrm{cm}$). See text and Appendix~\ref{realization} for details.
  }
  \label{fig:139freezein}
\end{figure}

\begin{figure}[h!]
\centering
  \includegraphics[width=0.9\linewidth]{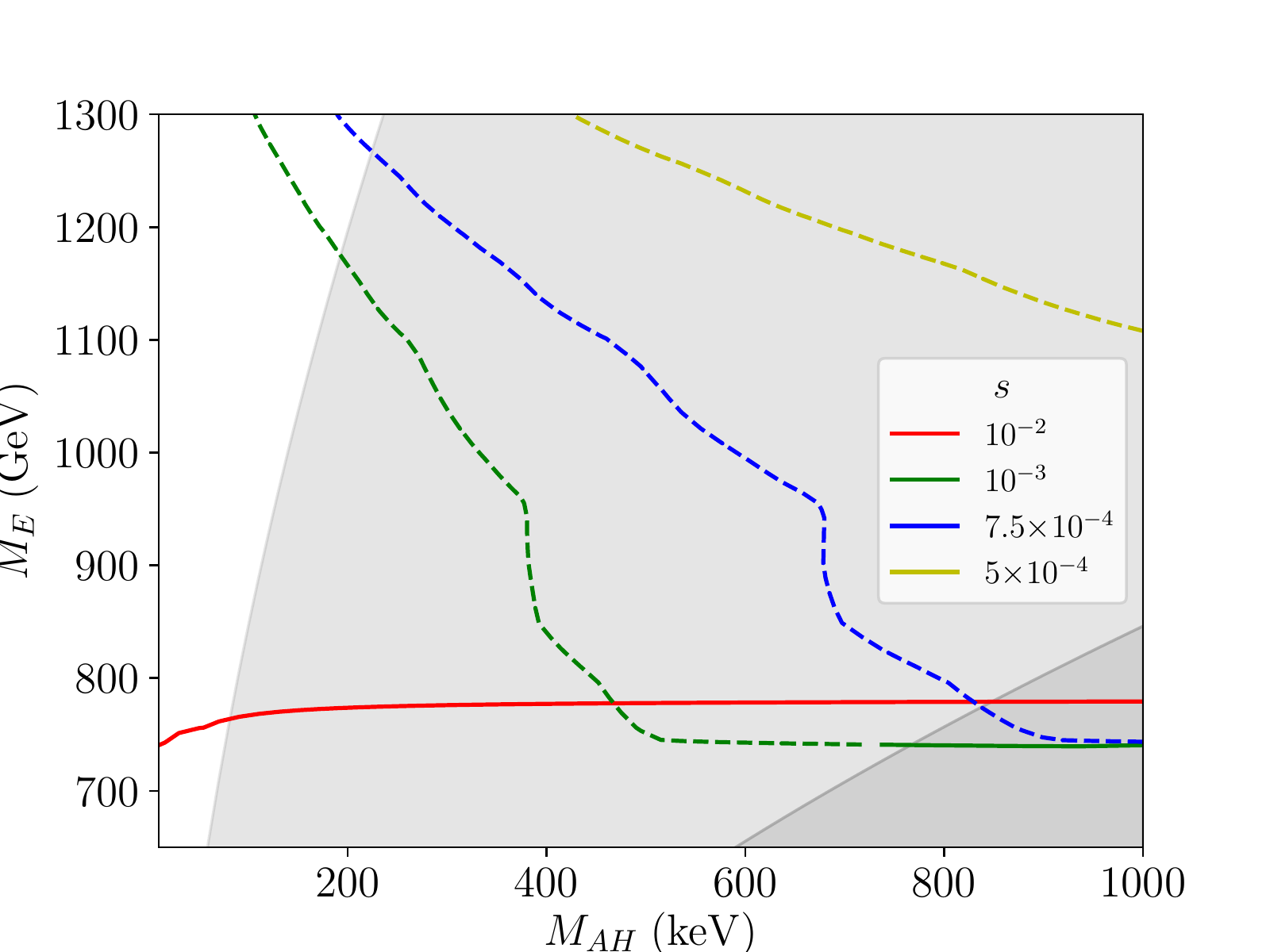}
  \caption{Contours of collider limits, for the LHC with
    $\sqrt{s}=13\;\mathrm{TeV}$ 
    and $\mathcal{L}=3\;\mathrm{ab}^{-1}$, on models that generate the
    observed DM relic abundance via the freeze-in mechanism for
    different values of the fermion mixing parameter $s$. The
    region below the curves is excluded by collider searches, either
    targeting SM decays (solid) or decays into missing energy
    (dashed). The light (dark) shaded region correspond to an
    $E\rightarrow A_H \ell$ decay length $\geq 1\;\mathrm{cm}$ ($\geq
    5\;\mathrm{cm}$). See text and Appendix~\ref{realization} for details.
  }
  \label{fig:3abfreezein}
\end{figure}

We present these results in Figures~\ref{fig:139freezein}
and~\ref{fig:3abfreezein} for the LHC analysis at 
$\sqrt{s}=13\;\mathrm{TeV}$ and
$\mathcal{L} = 139 \; \mathrm{fb}^{-1}$ and $\mathcal{L} = 3 \; \mathrm{ab}^{-1}$
respectively. The region below the curves can be excluded by collider
searches. In the region above the curve, for that fixed value of $s$,
all the values of $M_E$ and $M_{A_H}$ are experimentally allowed and
can provide the correct DM relic abundance. For each curve, we display
it as a solid or dashed curve depending on whether the most
constraining analysis is the one which is most sensitive to the decays into SM particles or
into missing energy, respectively.
This is relevant since, in
order to use the collider bounds we obtained, $E\rightarrow A_H \ell $
must be a prompt decay for the missing energy search whereas
$E\rightarrow Z\ell$ is the most important channel to be prompt in the
SM decays analysis. 

In Eq. (\ref{eq:zprompt}) we show the minimum value $s$ must take so
that $E\rightarrow Z\ell$ is prompt. For $E\rightarrow A_H \ell $ the
value of $g_H$ (fixed for each mass point) is going to determine
whether it is a prompt decay mode. In Figures~\ref{fig:139freezein}
and \ref{fig:3abfreezein} we show, in shaded gray, the regions in
which the $E\rightarrow A_H \ell$ decay length is
$\geq 1\;\mathrm{cm}$ (light gray) or $\geq 5\;\mathrm{cm}$ (dark gray).
This is relevant for the dashed part of the curves and shows that, for
smaller values of $s$, the limits on $M_E$ might be significantly
weaker. A more detailed analysis, which is beyond the scope of the
present work, targeting displaced vertices has the potential to
significantly probe the allowed region of parameter space in this
class of models.


\section{Conclusions \label{conclusions}}

New vector-like leptons are quite common in extensions of the Standard
Model. In
minimal extensions, with no further new particles or anomalous
couplings, their decays are governed by their mixing with the Standard
Model
leptons, which is strongly constrained by electroweak precision
data. These constraints eliminate the possibility of substantial
single production, leaving Drell-Yan pair production as their dominant
production mechanism. Realistic new physics models are, however, usually
far from minimal and the new particles present in the spectrum can
have a significant impact on the phenomenology of these new
leptons. New stable particles allow the possibility of a decay of the
vector-like lepton into a Standard Model charged lepton and missing
energy. Such a signature has 
been only experimentally searched for in the context of supersymmetric
models with slepton pair production decaying into leptons and
neutralinos. From the information given in the experimental analyses
it is difficult to directly translate the corresponding bounds to the
vector-like lepton
case, despite the fact that this signature is well motivated by
natural models like the Little Higgs models with T
parity. Furthermore, the case in which the new 
lepton can simultaneously decay into Standard Model particles and into
a Standard Model charged 
lepton and missing energy has been never considered before. This
possibility is however also well motivated as it naturally appears in
models of feebly interacting dark matter models in which the dark matter relic
abundance is generated via 
the freeze-in mechanism.

In order to fill this gap we have considered the possibility of a new
charge -1 vector-like lepton that can decay, with arbitrary branching
ratios into a 
Standard Model lepton together with a $Z$, $H$, $W$ or missing energy,
represented 
by a dark photon $A_H$, which
is assumed to be stable at detector scales.
We have then considered the most relevant LHC analyses probing such a
model and, after carefully validating our implementation of the
analyses, we have computed the current and future constraints that
hadron colliders can place on new vector-like leptons with these exotic decays.
Our results, represented as mass limits as functions of
$\mathrm{BR}(E\to A_H \ell)$ and
$\mathrm{BR}(E\to Z \ell)$ are provided in
Figures~\ref{fig:triangle_139}-\ref{fig:triangle_FCC} 
for current data at the LHC, the HL-LHC, the HE-LHC and the
100 TeV hh-FCC, respectively. This is one of the main results of our
work, as it provides the experimental limits from current and future
hadron colliders on a large number of models of vector-like leptons
with exotic decays. 

We have also considered the interesting possibility that the dark
photon, $A_H$, is not only stable at detector scales but also at
cosmological scales. It can then be a good dark matter candidate and we have
explored the interplay between the dark photon and the vector-like
lepton to provide
a successful explanation for the observed dark matter relic abundance.
After showing that the standard freeze-out mechanism  presents tension between the generation of the dark matter relic abundance and
limits from direct detection experiments, leaving only a relatively
small region of viable parameter space, we consider the case of near
degeneracy between the vector-like lepton and the dark photon. This leads to a
successful generation of dark matter via co-annihilation, compatible with all
current experimental limits. The relevant region of parameter space is
complementary to collider searches, as the compressed spectrum
significantly deteriorates the collider reach. The possibility of
specific searches that target these compressed spectra models becomes
a very interesting probe of the model in this regime.

Finally, we have considered the case in which the dark photon is very
light and feebly interacting, realizing the freeze-in mechanism. We
have shown that in this case collider searches are very complementary
to dark matter probes and we have found that models compatible with
current dark matter
phenomenology can be easily tested in current or future hadron
colliders.

\section*{Acknowledgments}

We are grateful to N. Castro, M. Chala, M. Ramos and T. Vale for useful
comments. This work has been supported in part by the Ministry of
Science, Innovation and Universities (PID2019-106087GB-C22) and  
	and by the Junta de Andaluc{\'\i}a grants FQM 101,
        SOMM17/6104/UGR, A-FQM-211-UGR18 and P18-FR-4314 (FEDER). GG acknowledges support by LIP (FCT, COMPETE2020-Portugal2020, FEDER, POCI-01-
0145-FEDER-007334) as well as by FCT under project CERN/FIS-PAR/0024/2019 and under the grant SFRH/BD/144244/2019.
        
        \appendix

\section{Explicit realization\label{realization}}

We describe in this appendix an explicit realization of the framework
used in this work. Rather than aiming at full generality we focus on a
minimal model capable of generating the range of branching
ratios we can be sensitive to at the LHC and future colliders. The
explicit realization we describe here is well motivated as a good
candidate for feebly interacting
DM~\cite{Delaunay:2020vdb}.~\footnote{Indeed our model corresponds 
to the one in~\cite{Delaunay:2020vdb} with the following replacements:
$M_0 \to M_E, \Lambda_1 \to x_E, V \to \omega/\sqrt{2},s\to \theta_R$.}
The model has an $SU(3)_C\times SU(2)_L\times
U(1)_Y \times U(1)_H$ gauge symmetry. The matter fields consist of the
SM particles, which are all neutral under $U(1)_H$, a new vector-like
lepton with the following quantum numbers, with notation $(SU(3)_C,SU(2)_L)_{U(1)_Y,U(1)_H}$,
\begin{equation}
  E^{(0)}_{L,R} \sim (1,1)_{-1,1},
\end{equation}
and a complex
scalar
\begin{equation}
\Phi\sim(1,1)_{0,1}.
\end{equation}
At the renormalizable level we can write the following Lagrangian
\begin{equation}
  \mathcal{L}=\mathcal{L}_{\mathrm{SM}} - \frac{1}{4} F_H^{\mu\nu}
  F_{H\,\mu\nu}+|D_\mu \Phi|^2 - V(\Phi)+ \bar{E}^{(0)}(\mathrm{i}
  \cancel{D} - M_0) E^{(0)}
  -\Lambda_1 (\bar{E}^{(0)}_L \Phi e^{(0)}_R + \mathrm{h.c.}) + \ldots,
\end{equation}
where $V(\Phi)$ is a suitable potential to spontaneously break
$U(1)_H$ and to make the physical Higgs scalar of such breaking much
heavier than all the 
other fields in the spectrum so that we can effectively neglect it.
For simplicity we have assumed
that kinetic
mixing between the two abelian groups is negligible~\footnote{The
order of magnitude expectation for kinetic mixing~\cite{Holdom:1985ag}
is small enough to be negligible for most of the parameter space and
also well within the experimental limits~\cite{Chun:2010ve}. For
values of $s$ on the smaller side a small extra suppression might be
needed~\cite{Gherghetta:2019coi}.} 
and that the VLL only couples to one of the
SM RH charged leptons, taken to be the electron here, denoted by
$e^{(0)}_R$ but it could equally well be the muon or tau, in the basis
of diagonal 
charged lepton Yukawa couplings.
Hereafter we suppress all terms in the Lagrangian that are not
relevant for our discussion.
The covariant
derivative for the new fields reads
\begin{equation}
  D_\mu = \partial_\mu - \mathrm{i} g_H A_{H\,\mu},
\label{eq:gh}
\end{equation}
where we have used $Q_H=1$.

Once $U(1)_H$ is spontaneously broken, the corresponding gauge boson, $A_H$,
acquires a mass
\begin{equation}
  M_{A_H}=\sqrt{2} g_H V,
\end{equation}
where we have denoted $V\equiv \langle \Phi \rangle$ the vacuum
expectation value (vev)
of $\Phi$,  
and $e^{(0)}_R$ and $E^{(0)}_R$ mix
\begin{equation}
  \mathcal{L}=- \bar{E}^{(0)}_L(\Lambda_1 V e^{(0)}_R + M_0
  E^{(0)}_R) + \mathrm{h.c.} + \ldots~.
\end{equation}
This mixing can be rotated away (thus defining the SM RH charged
lepton) via the following unitary rotation
\begin{equation}
  \begin{pmatrix} e_R^{(0)} \\ E^{(0)}_R \end{pmatrix}
  = \begin{pmatrix} c & s \\ -s & c \end{pmatrix}
  \begin{pmatrix} e_R \\ E_R \end{pmatrix},
\label{eq:s}
\end{equation}
where 
\begin{equation}
  s\equiv \frac{\Lambda_1 V}{M},
  \quad
  c \equiv \frac{M_0}{M},
  \quad
  M\equiv \sqrt{M_0^2 + \Lambda_1^2 V^2}.
  \end{equation}
Denoting $E^{(0)}_R \equiv E_L$ we have the SM extended with a VLL
singlet with hypercharge $-1$ and the following mass Lagrangian for
the charged leptons
\begin{equation}
  \mathcal{L}=
  \begin{pmatrix}
    \bar{e}_L & \bar{E}_L
  \end{pmatrix}
  \begin{pmatrix}
    m & m^\prime \\ 0 & M
  \end{pmatrix}
  \begin{pmatrix}
    e_R \\ E_R
  \end{pmatrix}
  + \ldots,\label{massmatrix:VLL}
\end{equation}
where $m$
and $m^\prime$ are generated after EWSB and satisfy
\begin{equation}
  \frac{m^\prime}{m}=\frac{s}{c},\label{mpoverm}
\end{equation}
and a new neutral heavy gauge boson with couplings
\begin{equation}
  \mathcal{L}=
  g_H A_H^\mu \begin{pmatrix} \bar{e} & \bar{E} \end{pmatrix}
  \gamma_\mu \left[
    \begin{pmatrix} 0 & 0 \\ 0 & 1 \end{pmatrix} P_L +
    \begin{pmatrix} s^2 & -sc \\ -sc & c^2 \end{pmatrix} P_R
    \right] \begin{pmatrix} e \\ E \end{pmatrix}  + \ldots~.
\end{equation}

The effect of mixing with extra vector-like fermions is well
known~\cite{delAguila:1982fs}. The physical basis is obtained by
diagonalizing the mass matrix in~(\ref{massmatrix:VLL}) via a
bi-unitary rotation
\begin{equation}
  \begin{pmatrix} e_\chi \\ E_\chi \end{pmatrix}
  \longrightarrow \begin{pmatrix} c_\chi & s_\chi \\ -s_\chi &
    c_\chi \end{pmatrix}
  \begin{pmatrix} e_\chi \\ E_\chi \end{pmatrix},
  \end{equation}
  where $\chi=L,R$ denotes the chirality and, in the $m^\prime \ll M$
  limit that we will be interested in we have
  \begin{equation}
    s_L=
    \frac{m^\prime}{M}+ \ldots
    , \quad s_R = \frac{m m^\prime}{M^2}+\ldots,
  \end{equation}
  where the dots denote higher orders in $m^\prime/M\ll 1$. The
  corresponding masses are
  \begin{equation}
    m_e = m \left(1- \frac{m^{\prime\,2}}{2 M^2} +\ldots\right)
    \approx m,
    \quad
    M_E= M \left(1+ \frac{m^{\prime\,2}}{2 M^2} +\ldots\right)
    \approx M.
    \end{equation}
In this physical basis, the coupling of
fermions to the electroweak gauge bosons, $Z$, $W$, the Higgs boson,
$H$, and the heavy photon, $A_H$, can be written as follows
\begin{align}
\mathcal{L}^Z =& \frac{g}{2c_W}Z_\mu \bar{\psi}^i_Q \gamma^\mu
        [X^{QL}_{ij} P_L +X^{QR}_{ij} P_R - 2 s_W^2 Q \delta_{ij} ] 
        \psi^j_Q,
        \nonumber \\ 
\mathcal{L}^W =& \frac{g}{\sqrt{2}} W^+_\mu\bar{\psi}^i_Q \gamma^\mu
        [V^{QL}_{ij} P_L +V^{QR}_{ij} P_R] \psi^j_{Q-1} 
        + \mathrm{h.c.},
        \nonumber \\ 
\mathcal{L}^H =& -\frac{H}{\sqrt{2}} \bar{\psi}^i_Q Y^Q_{ij} P_R
\psi^j_Q + \mathrm{h.c.},
\nonumber \\
\mathcal{L}^{A_H} =& g_H A_{H\,\mu}\bar{\psi}^i_Q \gamma^\mu
[Z^{QL}_{ij} P_L + Z^{QR}_{ij} P_R] \psi^j_Q,
\end{align}
where $\psi^i_Q$ is a fermion of
electric charge $Q$, $g$ is the $SU(2)_L$ gauge coupling, $c_W$ is the
cosine of the weak angle and $i,j$ are 
flavor indices. The relevant couplings are, to leading order in the
small $m^\prime/M$ expansion parameter,
\begin{align}
  X^{-1}_L \approx & \begin{pmatrix} -1 & -\frac{m^\prime}{M}
    \\ -\frac{m^\prime}{M} & -\frac{m^{\prime\,2}}{M^2} \end{pmatrix},
  \quad
  X^{-1}_R = (0),
  \nonumber \\
  W^0_L \approx& \begin{pmatrix} U_{i1}  & U_{i,1} \frac{m^\prime}{M} \end{pmatrix},
  \quad W^0_R \approx (0),
  \nonumber \\
  v Y^{-1} \approx& \begin{pmatrix} m & m^\prime \\ m
    \frac{m^\prime}{M} & \frac{m^{\prime\,2}}{M^2} \end{pmatrix},
  \nonumber \\
  Z^{-1}_L \approx & \begin{pmatrix} \frac{m^{\prime\,2}}{M^2} &
    -\frac{m^\prime}{M} \\ -\frac{m^\prime}{M} & 1 \end{pmatrix},
  \quad
  Z^{-1}_R \approx \begin{pmatrix}
    s^2 + 2 s c \frac{m m^\prime}{M^2} & -sc+(s^2-c^2)\frac{m m^\prime}{M^2}
\\ -sc+(s^2-c^2)\frac{m m^\prime}{M^2} & c^2 -2 s c \frac{m
  m^\prime}{M^2}
  \end{pmatrix},
\end{align}
where $v\approx 174$ is the Higgs vev,
$i$ denotes the neutrino flavor and we have suppressed all input
that is not directly relevant for our purposes.

In order to realize our scenario we consider the limit
$M_{A_H} \ll M$, so that $E$ can decay into
$Ze$, $He$, $W \nu$ 
and $A_H e$ and $A_H$ can decay into $\bar{e}e$ provided $M_{A_H}>2
m_e$.
The corresponding decay widths are
\begin{align}
&\sum_{i=1}^3\Gamma(E\to W \nu_i)\approx 
\sum_i \frac{g^2}{64 \pi} \Big[(V^{0}_L)^2_{iE} +(V^{0}_R)^2_{iE} \Big]
\frac{M_E^3}{m_W^2}
\approx 
\frac{g^2s^2}{64 \pi c^2} 
\frac{m_e^{2}M_E}{m_W^2},
\\
&\Gamma(E\to Z e)\approx 
\frac{g^2}{128 \pi c_W^2} \Big[(X^{-1}_L)^2_{eE} +(X^{-1}_R)^2_{eE} \Big]
\frac{M_E^3}{m_Z^2}
\approx
\frac{g^2s^2}{128 \pi c_W^2 c^2} 
\frac{m_e^{2}M_E}{m_Z^2},
\\
&\Gamma(E\to H e) \approx
\frac{1}{64\pi} \Big[ |(Y^{-1})_{eE}|^2 + |(Y^{-1})_{Ee}|^2\Big] M_E
\left( 1 -2\frac{m_H^2}{M_E^2} \right)
\nonumber \\
&\phantom{\Gamma(E\to H e)}
\approx
\frac{s^2}{64 \pi c^2} \frac{m_e^{2} M_E}{v^2}
\left( 1 -2\frac{m_H^2}{M_E^2} \right),
\\
&\Gamma(E\to A_H e)
\approx 
\frac{g_H^2}{32 \pi} \Big[(Z^{-1}_L)^2_{eE} +(Z^{-1}_R)^2_{eE} \Big]
\frac{M_E^3}{M_{A_H}^2}
\nonumber \\
&\phantom{\Gamma(E\to A_H e)}\approx
\frac{g_H^2}{32 \pi} \left[\frac{m^{\prime\,2}}{M_E^2} +\left(-sc+(s^2-c^2)
  \frac{m_e m^\prime}{M_E^2}\right)^2 \right]
\frac{M_E^3}{M_{A_H}^2}
\approx
\frac{g_H^2 s^2 c^2}{32 \pi} \frac{M_E^3}{M_{A_H}^2},
\end{align}
where we have shown the leading terms in the $x/M_E$, with
$x=m_e, m_Z, m_W, m_H$, except for $m_H$, for which the subleading
term is relevant for low values of $M_E$. Using the properties
\begin{equation}
  m_W= \frac{g v}{\sqrt{2}} = c_W m_Z,
\end{equation}
we recover the standard $2:1:1$ decay pattern into $W$, $Z$ and $H$
for large values of $M_E$.
Finally, assuming $M_{AH} \gg 2 m_e$ we have
\begin{align}
  \Gamma(A_H \to e^+ e^-) \approx&
  \frac{g_H^2}{24 \pi} \left[
    (Z^{-1}_L)_{ee}^2 +(Z^{-1}_R)_{ee}^2\right] M_{A_H}
  \nonumber \\
  \approx &
  \frac{g_H^2}{24 \pi} \left[
    \left( \frac{m^{\prime \, 2}}{M_E^2} \right)^2 +\left(
    s^2+2 s c \frac{m_e m^\prime}{M_E^2} \right)^2\right] M_{A_H}
  \approx \frac{g_H^2 s^4}{24\pi} M_{A_H}.
  \end{align}
In order to realize our framework we need $A_H$ to be stable at
detector scales, $E$ to decay promptly, and the branching ratios of
$E$ decaying into $A_H$ and the SM bosons to be of similar
order. Assuming a decay length larger than $\sim 10 \;\mathrm{m}$ for $A_H$ and
smaller than $10^{-2} \;\mathrm{m}$ for $E$, these conditions translate into
\begin{align}
&\Gamma(A_H \to e^+ e^-) \lesssim 2\times 10^{-19}~\mathrm{GeV},\quad
\mbox{(Invisible $A_H$)},
\\
&\Gamma(E \to Z e, A_H e) \gtrsim 2\times 10^{-16}~\mathrm{GeV},\quad
\mbox{(Prompt $E$ decays)}.
\end{align}
Using the expressions above we can find, for each value of $M_E$ and
$M_{A_H}$, the values of $g_H$ and $s$ that satisfy these conditions. 
Indeed, requiring prompt $E\to Z e$ decays gives lower bound on $s$
\begin{equation}
  \frac{s}{c} \gtrsim \left[ \frac{128\pi c_W^2}{g^2}
    \frac{m_Z^2}{m_e^2 M_E} 2\times 10^{-16} \mbox{
      GeV}\right]^{\frac{1}{2}}
  = \left \{
  \begin{array}{l}
    3.1\times 10^{-3} \sqrt{\frac{500\mbox{ GeV}}{M_E}}, \mbox{
      electron},
    \\
    1.5\times 10^{-5} \sqrt{\frac{500\mbox{ GeV}}{M_E}}, \mbox{
      muon}.
  \end{array}
  \right.
\label{eq:zprompt}
  \end{equation}
Requiring that $A_H$ decays invisibly and the decay $E\to A_H e$ is
prompt provides in turn an upper limit on $s$
\begin{equation}
  \frac{\Gamma(A_H \to e^+ e^-)}{\Gamma(E \to A_H e)} \leq 10^{-3}
  \Rightarrow
  \frac{s}{c} \lesssim 2.7 \times 10^{-2} \left(\frac{M_E}{M_{A_H}}\right)^{\frac{3}{2}}.
\end{equation}
This upper limit is of the same order of magnitude as the one from
electroweak precision data~\cite{deBlas:2013gla}.
Note that for the values of $M_E$ we are sensitive to, unless
$M_{A_H}$ is very close to $M_E$, the two limits are always compatible.
Provided $s$ is fixed in the allowed range, we can fix a minimum value
of $g_H$ by requiring $E\to A_H e$ to be prompt
\begin{equation}
  g_H \gtrsim \left[
\frac{32 \pi}{s^2 c^2} \frac{M_{A_H}^2}{M_E^3} 2 \times 10^{-16} GeV 
\right]^{\frac{1}{2}}
  \approx
  \frac{1.3 \times 10^{-9}}{s c} \frac{M_{A_H}}{100\mbox{ GeV}}
  \left[\frac{500\mbox{ GeV}}{M_E}\right]^{\frac{3}{2}},
\end{equation}
and a maximum one by requiring $A_H$ to be stable at detector scales
\begin{equation}
  g_H \lesssim \left[\frac{24 \pi}{s^4} \frac{2\times
      10^{-19}}{M_{A_H}}\right]^{\frac{1}{2}} \approx
  \frac{  4 \times 10^{-10}}{s^2} \left[ \frac{100\mbox{
      GeV}}{M_{A_H}}\right]^{\frac{1}{2}}.
  \end{equation}
Once $s$ and $g_H$ are fixed within the allowed values we have fixed
the relative decay of $E$ into $A_H$ and $Z$
\begin{equation}
  \mathcal{R}\equiv \frac{\Gamma(E\to Z e)}{\Gamma(E \to A_H e)}
\approx
\frac{g^2}{4 c_W^2 c^4 g_H^2}
\frac{M_{A_H}^2}{m_Z^2}\frac{m_e^2}{M_E^2}
\approx
\frac{1.65 \times 10^{-13}}{g_H^2 c^4}
\left(\frac{M_{A_H}}{100\mbox{ GeV}}\right)^2 \left(\frac{500\mbox{ GeV}}{M_E}\right)^2.
\end{equation}
Using the minimum and maximum values of $g_H$ we get
\begin{align}
  10^{6} \frac{s^4}{c^4}
\left(\frac{M_{A_H}}{100\mbox{ GeV}}\right)^3
  \left(\frac{500\mbox{GeV}}{M_E}\right)^2
  \lesssim \mathcal{R} \lesssim
  10^{5} \frac{s^2}{c^2} \left(\frac{M_E}{500 \mbox{ GeV}}\right).
\end{align}
As an example, fixing $M_E=500\mbox{ GeV}$ and $M_{A_H}=100\mbox{
  GeV}$ we have
\begin{equation}
  3.1\times 10^{-3} \lesssim \frac{s}{c} \lesssim 0.094.
\end{equation}
Fixing for instance $s=0.01$ we now have
\begin{equation}
  1.3\times 10^{-7} \lesssim g_H \lesssim 4\times 10^{-6},
\end{equation}
and
\begin{equation}
  10^{-2} \lesssim \mathcal{R} \lesssim 10.
\end{equation}
Considering the muon instead of the electron increases $\mathcal{R}$ by
a factor $(m_\mu/m_e)^2\approx 4.4\times 10^4$ and reduces the lower
limit of $s/c$ by a factor $m_e/m_\mu \approx 4.8\times 10^{-3}$.

\bibliographystyle{JHEP}
\bibliography{VLLexoticdecays}
	
\end{document}